\documentclass[12pt]{article} 
\usepackage{amsmath,amssymb}

\usepackage{pst-all,citesort}    
\usepackage[round,authoryear]{natbib} 

\def\cocoa{{\hbox{\rm C\kern-.13em o\kern-.07em C\kern-.13em o\kern-.15em A}}}
\newcounter{definicion}
\newcounter{ejemplo}

\newtheorem{theorem}{Theorem}

\newtheorem{conjecture}[theorem]{Conjecture}

\newtheorem{definition}[definicion]{Definition}
\newtheorem{example}[ejemplo]{Example}

\newtheorem{lemma}[theorem]{Lemma}

\newtheorem{remark}[theorem]{Remark}

\newenvironment{proof}[1][Proof]{\textbf{#1.} }{\ \rule{0.5em}{0.5em}}

\title{Minimal average degree aberration and the state polytope for experimental designs}

\author{Yael Berstein\footnote{Technion, Israel Institute of Technology, Haifa 32000, Israel},
Hugo Maruri-Aguilar\footnote{Department of Statistics, London School
of Economics, London WC2A 2AE, UK}, Shmuel Onn$^*$, \\Eva
Riccomagno\footnote{Dipartimento di Matematica, Universit\`a di
Genova,
  Genova 16146, Italy}, Henry Wynn$^\dagger$}

\begin{document}
\maketitle

\begin{abstract}

For a particular experimental design, there is interest in finding
which polynomial models can be identified in the usual regression
set up. The algebraic methods based on Gr\"{o}bner bases
provide a systematic way of doing this. The algebraic method does
not in general produce all estimable models but it can be shown that
it yields models which have minimal average degree in a well-defined
sense and in both a weighted and unweighted version. This provides
an alternative measure to that based on ``aberration" and moreover
is applicable to any experimental design.  A simple algorithm is
given and bounds are derived for the criteria, which may be used to
give asymptotic Nyquist-like estimability rates as model and sample
sizes increase.

\end{abstract}



\section{Introduction}

It is of considerable value to represent an experimental design as
the solution of a set of polynomial equations. In the terminology of
algebraic geometry a design is a zero dimensional variety and the
corresponding ideal comprising all polynomials which are zero on
every design point is called an ``ideal of points". Pistone \& Wynn
(1996) first used explicit methods from algebraic geometry and in
particular introduced Gr\"obner bases into designs. Issues to do
with identifiability of polynomial regression models, or
interpolators, can be translated into problems about such varieties
and ideals, see \cite{PRW2000}.

The purpose of this paper is to introduce the notion of linear
aberration of a polynomial model. Linear aberration is defined only
for polynomial models, which are used routinely in statistical
literature. A polynomial model with low order terms has low
aberration, thus engaging low aberration with the standard practice
of preferring polynomial models with low order terms. The preference
for models with low order terms has been acknowledged in recent
papers, see \cite{LLY2003} and \cite{BY2006}, although they do not
refer to linear aberration.

Let  $\alpha = (\alpha_1,\ldots, \alpha_d)$ be a nonnegative
$d$-dimensional integer multi-index. A monomial in the
indeterminates $x_1,\ldots,x_d$ is the power product $x^{\alpha} =
x_1^{\alpha_1}\cdots x_1^{\alpha_d}$.
A {\em model basis} is a collection of distinct monomials
$\{x^{\alpha}, \alpha \in L\}$, where $L$ is a finite set of
multi-indices. By combining linearly monomials in $L$ we form
polynomials:
$$\eta_L(x) = \sum_{\alpha \in L} \theta_{\alpha} x^{\alpha},$$
where $\theta_{\alpha}$ are real coefficients. The polynomial
$\eta_L(x)$ is a candidate for
interpolation or statistical modelling.

This paper is concerned with the following concepts.
\begin{definition}\label{def_ab} Let $L$ be a model basis and let
$w= (w_1, \ldots, w_d)$ be a collection of non-negative weights with
$\sum_{i=1}^d{w_i}=1$. We define the weighted linear aberration of
$L$ as
\begin{equation*}\label{ec_Abdef}A(w,L) =
\frac{1}{n}\sum_{(\alpha_1,\ldots,\alpha_d)\in
L}\sum_{i=1}^dw_i\alpha_i,
\end{equation*}
where $n$ is the number of elements in $L$.
\end{definition}

We are interested in studying aberration for models identifiable by
an experimental design and along this paper we compare models and
designs of the same size $n$.

\begin{definition} An experimental design $D$, of sample size
$n=|D|$, is a set of points in ${\mathbb R}^d$.
\end{definition}

We say that a model basis $L$ with cardinality $|L|=n$ is
identifiable by $D$ if the design model matrix $X=[x^\alpha]_{x\in
D,\alpha\in L}$ is invertible.

The term aberration is used to acknowledge the work on ``minimum
aberration" for regular fractional factorial designs of Wu and
others, see \cite{FH1980} and \cite{WW2002}. For fractional
factorial designs, the notion of estimation capacity is related to
the ability of a design to identify models of low degree, see
\cite{CM1998} and \cite{CC2004}.
We do not make a direct mathematical comparison with that work but
simply point to a common motivation.


In Section \ref{sec_GB} we review the basic ideas on algebraic
identifiability. The search for identifiable models is driven by a
divisibility condition, which makes the search problem tractable. We
then introduce the \textit{state polytope}, whose vertices
correspond to the models identified using the algebra. In Section
\ref{sec_Ab} we study aberration. The basic ideas on aberration are
closely linked with the algebraic work on corner cut models and
state polytopes in \cite{OS1999}. We are specially interested in
obtaining minimal values for aberration for which we establish upper
and lower bounds. An approximate approach to minimal aberration is
discussed. In Section \ref{sec_Ex} we discuss various examples. In
Section \ref{sec_Di} we discuss possible extensions of the theory
and, by example, a connection with the notion of aberration by Wu
and others is discussed.


\section{The G-basis method and the state polytope}\label{sec_GB}
The aberation $A(w,L)$ has remarkable connections with the algebraic
method in experimental design introduced by Pistone and Wynn
\cite{PW1996} and developed in the monograph \cite{PRW2000} and the
joint work of Onn and Sturmfels \cite{OS1999}. In this section we
present the basic ideas on identifiability using algebraic
techniques.

Let the set of all monomials in $d$ indeterminates be
$T^d=\{x^\alpha, \alpha\in\mathbb Z_{\geq 0}^d\}$, where $\mathbb
Z_{\geq 0}$ is the set of non-negative integers and $\mathbb Z_{\geq
0}^d$ is the set of all vectors in $d$ dimensions and with entries
in $\mathbb Z_{\geq 0}$. A polynomial is a finite linear combination
of monomials in $T^d$ with real coefficients. The set of all
polynomials is denoted as $\mathbb R[x_1,\ldots,x_d]$. It has the
structure of a ring with the usual operations of sum and product of
polynomials.

A term ordering $\succ$ on $\mathbb R[x_1,\ldots,x_d]$ is a total
ordering on $T^d$ such that i) $x^\alpha\succ1$ for all $x^\alpha\in
T^d$, $\alpha\neq(0,\ldots,0)$ and ii) for all
$x^\alpha,x^\beta,x^\gamma\in T^d$ if $x^\alpha\succ x^\beta$ then
$x^\alpha x^\gamma\succ x^\beta x^\gamma$. The leading term of a
polynomial is the largest term with non-zero coefficient with
respect to $\succ$. For a polynomial $f\in\mathbb
R[x_1,\ldots,x_d]$, we write its leading term as $\operatorname{
LT}_\succ(f)$.

A partial order on $T^d$ is defined by a vector $w\in\mathbb R_{\geq
0}^d$ as $x^\alpha\succeq_w x^\beta$ if $w^T\alpha\geq w^T\beta$,
where $x^\alpha,x^\beta\in T^d$ and $w^T$ is the transposed vector
of $w$. Under some conditions on $w$ (see \cite{BOT2003,CLO1996})
this defines a term order. Given a term order $\succ$, there are $w$
such that $x^\alpha\succ x^\beta$ if and only if $x^\alpha\succeq_w
x^\beta$.

A design $D$, considered as a zero-dimensional variety gives rise to
a {\em design ideal}, $I(D)$, which is the set of all polynomials
which have zeros at all the points of $D$. We have that $I(D)\subset
\mathbb R[x_1,\ldots,x_d]$. The polynomial ideal $I$ is generated by
the set of polynomials $G=\{g_1,\ldots,g_s\}$ if
$I=\left\{\sum_{i=1}^s{f_ig_i}:f_i\in\mathbb
R[x_1,\ldots,x_d]\right\}$ and we write $I=\langle
g_1,\ldots,g_s\rangle$.

An important set of generators for the design ideal is the Gr\"obner
basis. Gr\"obner bases were introduced by Buchberger in \cite{B1966}
and they have become a powerful computational tool in many fields
\cite{CLO1996,CLO2005}. A Gr\"obner basis of $I(D)$ with respect to
a term order $\succ$ is a finite subset $G_\succ(D)\subset I(D)$
such that $\langle \operatorname{LT}_\succ(g):g\in
G_\succ(D)\rangle=\langle\operatorname{LT}_\succ(f):f\in I(D)
\rangle$. The computation of Gr\"obner bases is implemented in
standard computer programs such as \cocoa, Singular or Maple, see
\cite{CocoaSystem,GPS2005,Maple10}.

Two polynomials $f$ and $g$ in $\mathbb R[x_1,\ldots,x_d]$ are
equivalent with respect to $I(D)$ if the following conditions
hold:\begin{enumerate}\item[i)] $f-g\in I(D)$\item[ii)] $f(d)=g(d)$
for all $d\in D$\end{enumerate}
 Given a term ordering $\succ$, the quotient ring
$\mathbb R[x_1,\ldots,x_d]/I(D)$ has a unique $\mathbb R$-vector
space basis given by the monomials in $T^d$ that cannot be divided
by the leading terms of the polynomials in $G_\succ(D)$ for $I(D)$.
The monomial basis so obtained, or equivalently, the set of its
exponents $L=L(D,\succ)$, has a {\em staircase} (also {\em echelon},
{\em order ideal}) property: for $\alpha \in L$, if $\beta \leq
\alpha$ componentwise, then $\beta \in L$. Equivalently we say that
for any $x^\alpha\in L$, if $x^{\beta}$ divides $x^{\alpha}$ then
$x^{\beta} \in L$. We call bases which have a staircase structure
{\em staircase} models. The dimension of $\mathbb
R[x_1,\ldots,x_d]/I(D)$ as $\mathbb R$-vector space is $n$, see
\cite{PW1996}, i.e. the number of points in $D$ and of multi-indices
in $L$ is $n$.

For a given basis of the quotient ring with exponents in $L$ and a
set of real values (data) $Y_x, x \in D$, there exists a unique
interpolator $\eta_L(x)$ such that $Y_x=\eta_L(x),\; x \in D$. Other
non-saturated statistical sub-models can be constructed from subsets
of $L$, see \cite{PHRWPcasestudy} and \cite{P1987}.

\begin{definition}
The algebraic fan of $D$ is $\mathcal
 L_a(D)=\{L(D,\succ)$, where $\succ$ is a term ordering in
 $\mathbb R[x_1,\ldots,x_d]\}$. This is the collection of staircases
 $L(D,\succ)$ arising from a fixed design $D$ by varying all
 monomial orderings.
\end{definition}

The algebraic fan of a design was proposed by Caboara \textit{et
al}. \cite{CPRW1997}, constructing upon the algebraic fan of an
ideal of Mora and Robbiano \cite{MR1988}. Babson \textit{et al}.
\cite{BOT2003} proposed a polynomial time algorithm to compute
$\mathcal
 L_a(D)$. They compute an efficient set of weight
vectors and perform a change of basis which stems from the so-called
FGLM algorithm, see \cite{FGLM1993}. In Section
\ref{sec_compute_model} an algorithm is presented to identify a
model in the algebraic fan using a weight vector.

It is important to note that not all staircase models identified by
$D$ are in $\mathcal
 L_a(D)$. We denote the set
of all identifiable staircase models for a design $D$ as $\mathcal
L_s(D)$. In fact the algebraic fan is small relative to $\mathcal
L_s(D)$, that is $\mathcal L_a(D) \subseteq \mathcal L_s(D)$, see
Chapter 6 in the unpublished Ph.D. thesis by Maruri-Aguilar (2007)
and Section 4 in \cite{PRR2006}.

We now establish the link between the algebraic fan of a design and
the state polytope of the design ideal. For a model basis
$L=\{\alpha_1,\ldots,\alpha_n\},\alpha_i\in\mathbb Z_{\geq 0}^d$
define
$$\overline{\alpha}_L = \sum_{\alpha_i \in L} \alpha_i.$$
This vector appears in the definition of $A(w,L)$ and we can write
$A(w,L)=(w^T\overline{\alpha}_L)/n$. The set all such vectors over
$\mathcal L_a(D)$ gives the state polytope.
\begin{definition}
The state polytope $S(D)$ of a design $D$, or equivalently of the
design ideal $I(D)$ is the convex hull
$${\mathcal S(D)}:= \operatorname{conv}\left(\{\overline{\alpha}_L : L\mbox{ is a staircase in } \mathcal L_a(D)  \}\right).$$
\end{definition}

The following theorem \cite[Ch. 2]{S1995} summarizes the connection
between the state polytope and the set of models $\mathcal L_a(D)$,
i.e. the relation between a design and its algebraic fan.

\begin{theorem} 
Let $D$ be a design and let $S(D)$ be its state polytope. Then the
set of vertices of the state polytope of $D$ is in one to one
correspondence with the algebraic fan of $D$.
\end{theorem}

The state polytope does not only contain information concerning
models in the algebraic fan of a design, but it also provides
information about the term ordering vectors needed to construct it.
We recall that a $d$-dimensional polytope is a bounded subset of
$\mathbb R^d$, which corresponds to the solutions of a system of
linear inequalities. The \textit{normal cone} of a face of a
polytope is the relatively open cone of those vectors in $\mathbb
R^d$ uniquely minimised over the face of the polytope. The
\textit{normal fan} of a polytope is the collection of all the
normal cones of the polytope.

Two ordering vectors $w$ and $w'$ are said to be equivalent (modulo
$I(D)$) if \mbox{$L(D,\succ_w)=L(D,\succ_{w'})$}. The normal fan of
the state polytope partitions $\mathbb R^d_{\geq 0}$ into
equivalence classes of ordering vectors, see
\cite{BOT2003,FJT2005,S1995}. Indeed every vertex of $S(D)$
corresponds to a model in $\mathcal L_a(D)$. Moreover, the interior
of the normal cone of a vertex in $S(D)$ contains those vectors $w$
which correspond to the same equivalence class.

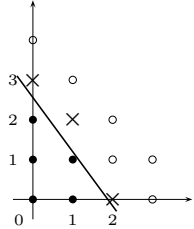
\begin{figure}[h]  
\begin{center}
\psset{unit=5.3mm,linewidth=0.8pt}
 \begin{pspicture}(-1,-1)(4,6)
     \put(0,0){}
\psset{linecolor=black,linewidth=0.25pt}
     \psline{->}(0,-0.5)(0,5.0)
     \psline{->}(-0.5,0)(4.0,0)
     \pscircle*[fillcolor=black](0,0){0.1}
     \pscircle*[fillcolor=black](1,0){0.1}
     \pscircle*[fillcolor=black](0,1){0.1}
     \pscircle*[fillcolor=black](1,1){0.1}
     \pscircle*[fillcolor=black](0,2){0.1}
     \pscircle[fillcolor=black](0,4){0.1}
     \pscircle[fillcolor=black](2,1){0.1}
     \pscircle[fillcolor=black](3,0){0.1}
     \pscircle[fillcolor=black](1,3){0.1}
     \pscircle[fillcolor=black](2,2){0.1}
     \pscircle[fillcolor=black](3,1){0.1}
          \psdot*[dotstyle=x,dotsize=0.35](2,0)
          \psdot*[dotstyle=x,dotsize=0.35](1,2)
          \psdot*[dotstyle=x,dotsize=0.35](0,3)

     \rput(1,-0.5){\tiny{$1$}} \rput(2,-0.5){\tiny{$2$}}
     \rput(-0.5,1){\tiny{$1$}} \rput(-0.5,2){\tiny{$2$}} \rput(-0.5,3){\tiny{$3$}}
     \rput(-0.35,-0.5){\tiny{$0$}}
     \psset{linewidth=0.6pt}
\psline(-0.4,3.1)(2.1,-0.3)
\end{pspicture}

\caption{Corner cut and separating hyperplane.} \label{fig_cc1}

\end{center}
\end{figure}

We motivate Theorem \ref{th_generic} below with a simple example.
The black dots in Figure \ref{fig_cc1} give a $5$ point design in
$2$ dimensions, $D$. They also give the set of exponents $L$
obtained for any term ordering, indeed the size of the algebraic fan
of $D$ is one. The crosses represent the exponents of the leading
terms of the Gr\"{o}bner basis: $(2,0),(1,2),(0,3)$. The line
separates the model exponents, $L$, from these leading terms. This
is an example of a {\em corner cut} model. Note that equivalently
the line separates $L$ from its complement in $\mathbb Z_{\geq
0}^2$.

\begin{definition}
A model $L$, of size $|L|=n$, is said to be a corner cut model if
there is a $(d-1)$ dimensional hyperplane separating $L$ from its
complement $\mathbb Z_{\geq 0}^d \setminus L$.
\end{definition}

Not all staircases are corner cuts, for example
$L=\{(0,0),(1,0),(0,1),(1,1)\}$ is a staircase that cannot be
separated by a hyperplane from its complement in $\mathbb Z_{\geq
0}^2$.

The set of exponents of a corner cut model is referred to as a
corner cut staircase or simply, as a corner cut. Corner cuts were
introduced by Onn and Sturmfels \cite{OS1999}. A generating function
for the number of bidimensional corner cuts is given in
\cite{CRST1999}, while the order of the cardinality of the set of
corner cuts is proven bounded by $(n\log n)^{d-1}$ in \cite{W2002}.
A special class of designs is composed with those designs that
identify all corner cut models of a given size.

\begin{definition}\label{def_generic}
A design $D\subset\mathbb R^d$ comprised of $n$ distinct points is
said to be generic if all corner cut models of size $n=|D|$ are
identifiable.
\end{definition}

A special polytope is constructed with the exponents for corner cut
models. It will be used to compute the algebraic fan of generic
designs.

\begin{definition}
The corner cut polytope is $CC(n,d):=\operatorname{conv}(\{
\overline{\alpha}_L : L$ is a corner cut staircase in $d$
dimensions and of size $n\})$. 
\end{definition}

For a discussion on the properties of bidimensional corner cut
polytopes see the paper by M\"uller \cite{IM2003}. The algebraic fan
of generic designs corresponds to the set of corner cut models, as
stated in the following theorem.

\begin{theorem} \label{th_generic}(Onn and Sturmfels, 1999)
Let $D\subset\mathbb R^d$ be a generic design with $n$ points. Then
\begin{enumerate}\item[i)] $S(D)=CC(n,d)$ and
\item[ii)] the algebraic fan of $D$ is the set of corner
cut models in $d$ dimensions and with $n$ elements.\end{enumerate}
\end{theorem}

We remark that the corner cut polytope is an invariant object for
the class of all the ideals generated by generic designs with the
same sample size $n$ and number of factors $d$ and all generic
designs have the same state polytope.


\section{Minimal linear aberration} \label{sec_Ab}
An important feature of the state polytope is that its vertices are
automatically ``lower" vertices in the sense of convexity. State
polytopes relate directly to models with minimal linear aberration.
In Section \ref{sec_compute_model} an algorithm to compute a models
of minimal aberration is presented.

\begin{theorem} \label{th_vertex_mini}
Given a design $D\subset\mathbb R^d$ with $n$ distinct points and a
weight vector $w\in\mathbb R_{>0}^d$, there is a least one vertex
$\alpha^*\in S(D)$ which minimises $A(w,L)$ over all identifiable
staircase models $\mathcal L_s(D)$, that is
$$\frac{1}{n}(w^T\alpha^*)=A(w,L^*)=\min_{L\in\mathcal L_s(D)}A(w,L)$$ for all
$L^*$ such that $\overline\alpha_{L^*}=\alpha^*$. Moreover, given a
vertex of $S(D)$, there is at least one $w^*\in\mathbb R_{>0}^d$
such that this vertex (model) minimizes $A(w,L)$, that is,
$$A(w^*,\overline L)=\min_{w\in\mathbb R_{>0}^d}A(w,L)$$ for $\bar L$ such that
$\overline\alpha_{\overline L}=\overline\alpha_L$.

\end{theorem}
\begin{proof}
First, for given $w$ we minimise $w^T\overline\alpha_L$ for
$L\in\mathcal L_a(D)$, which is a finite set, see \cite{MR1988}. The
$\overline\alpha_L$ for $L\in\mathcal L_a(D)$ are vertices of $S(D)$
by definition. Furthermore, because we restrict $L$ to the algebraic
fan of $D$ there cannot be three aligned $\overline\alpha_L$ in
$S(D)$, see \cite{S1995}. For the second claim, it is sufficient to
take a vector $w_L$ in the interior of a normal cone for
$\overline\alpha_L$. By definition, $A(w,L)$ is minimised for
vectors on the interior of the normal cone.
\end{proof}

Theorem \ref{th_exist} follows directly from Theorem
\ref{th_vertex_mini}.

\begin{theorem} \label{th_exist}
For every weight vector $w$ there is a design $D\subset\mathbb R^d$
which minimizes $A(w,L)$, among all designs with sample size $n$ and
identifiable staircases.
\end{theorem}
This is stated compactly as:
$$A^*(w,n)=\min_{D:|D|=n} \min_{L \in {\mathcal L}_a(D)} A(w,L)$$
is achieved for a generic design. That is, if a design is generic
then automatically its algebraic fan contains models of minimal
aberration.


\subsection{Computation of the minimal aberration
model}\label{sec_compute_model}

The model minimizing linear aberration can be found by the {\em
greedy algorithm}. Let $D$ be a design; let $w$ be a fixed weight
vector in $\mathbb R^d_{>0}$ and let $\Gamma$ be the following set
of potential exponents
$$\Gamma :=\left \{\alpha=(\alpha_1,\dots,\alpha_d)\in
\mathbb{Z}^d_{\geq 0} : \prod_{i=1}^d(\alpha_i+1)\leq n\right\}.$$
The set $\Gamma$ contains all staircase models with $n$ terms, see
\cite{BOT2003}. Now define the {\em weight} of $\alpha\in\Gamma$ to
be $\omega(\alpha):=\frac{1}{n}\sum_{i=1}^dw_i\alpha_i=(w^T\alpha)/n$. Order the vectors in
$\Gamma$ by their weight $\omega(\cdot)$ in increasing order, that
is, index them as $\alpha^1,\dots,\alpha^{|\Gamma|}$ such that
$\omega(\alpha^1)\leq \cdots \leq \omega (\alpha^{|\Gamma|})$, where
$|\Gamma|$ is the cardinality of $\Gamma$. Then the set
$L\subseteq\Gamma$ with the first $n$ terms of $\Gamma$ which are
identifiable by $D$ has minimum aberration.

The model basis $L$ is constructed by the following procedure:
initialize $L:=\emptyset$; while $|L|<n$, find $\alpha^i$ of
smallest index with respect to $\omega(\cdot)$ such that the column
vectors $d^\alpha$, $\alpha\in L\cup\{\alpha^i\},d\in D$, are
linearly independent; update $L:=L\cup\{\alpha^i\}$ and repeat until
$|L|=n$. We have the following theorem.

\begin{theorem} Let $D\subset\mathbb R^d$ be a design; let $w$ be a fixed
weight
vector with positive entries and let $L$ be the model basis
constructed by the greedy algorithm. Then $L$ belongs to the
algebraic fan of the design.
\end{theorem}

\begin{example}{\normalfont
Consider the design $D=\{(0,0),(1,0),(0,1),(-1,1)\}$ and the weight
vector $w=(4,1)$. The set of potential exponents, $\Gamma$ contains
$8$ elements, which are sorted out using the weight function
$\omega(\cdot)$ as
$$\begin{array}{lcccccccccl}
\Gamma&=\{&(0,0),&(0,1),&(0,2),&(0,3),&(1,0),&(1,1),&(2,0),&(3,0)&\}\\
n\omega(\cdot)&=&0&1&2&3&4&5&8&12&
\end{array}$$
 The
first $4$ elements in $\Gamma$ such that their design columns are
linearly independent are $L=\{(0,0),(0,1),(1,0),(0,1)\}$. Thus the
set $L$ of minimal linear aberration corresponds to the model with
terms $\{1,x_1,x_2,x_1x_2\}$.}\end{example}

\subsection{Examples}

We can compare different designs using aberration as long as they
have the same number of factors $d$ and the number of points $n$.
For a design $D$, the state polyhedron of $D$ is obtained by
(Minkowski) addition of $\mathbb R^d_{\geq 0}$ to the state polytope
$S(D)$, see \cite{BOT2003}. The state polyhedron yields the same
information as the state polytope. Indeed the normal fan of the
(negative) state polyhedron yields automatically the first orthant,
see Fukuda \textit{et al}. \cite{FJT2005}.


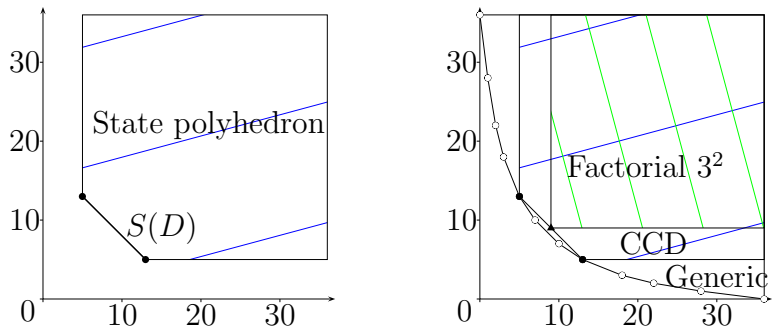
\begin{figure}[h]  
\begin{center}
\psset{unit=1.05mm,linewidth=0.5pt}
 \begin{pspicture}(-2,-2)(44,38)

 \psset{linewidth=0.25pt}
 \psline{<->}(0,37)(0,0)(37,0)
  \psline(0,0)(0,-0.4)
 \psline(10,0)(10,-0.4)
  \psline(20,0)(20,-0.4)
  \psline(30,0)(30,-0.4)
  \psline(0,10)(-0.4,10)
  \psline(0,20)(-0.4,20)
  \psline(0,30)(-0.4,30)
  \psline(0,0)(-0.4,0)
  \rput(10,-2.05){$10$}
  \rput(20,-2.05){$20$}
  \rput(30,-2.05){$30$}
  \rput(-2.3,10){$10$}
  \rput(-2.3,20){$20$}
  \rput(-2.3,30){$30$}
  \rput(-1.9,-1.55){$0$}

 \psset{linewidth=0.35pt,linecolor=black}
 \pspolygon[linewidth=0.25pt,fillstyle=vlines,hatchangle=105,hatchwidth=0.105pt,hatchsep=44pt,hatchcolor=blue]
(13,5)(5,13)(5,36)(36,36)(36,5) 
\psline[linewidth=0.5pt](13,5)(5,13)
 \psdots[dotstyle=*](13,5)(5,13)

\rput(21,22){State polyhedron} \rput(15,9){$S(D)$}


\end{pspicture}
 \begin{pspicture}(-10,-2)(38,38)

 \psset{linewidth=0.25pt}
 \psline{<->}(0,37)(0,0)(37,0)
  \psline(0,0)(0,-0.4)
 \psline(10,0)(10,-0.4)
  \psline(20,0)(20,-0.4)
  \psline(30,0)(30,-0.4)
  \psline(0,10)(-0.4,10)
  \psline(0,20)(-0.4,20)
  \psline(0,30)(-0.4,30)
  \psline(0,0)(-0.4,0)
  \rput(10,-2.05){$10$}
  \rput(20,-2.05){$20$}
  \rput(30,-2.05){$30$}
  \rput(-2.3,10){$10$}
  \rput(-2.3,20){$20$}
  \rput(-2.3,30){$30$}
  \rput(-1.9,-1.55){$0$}

 \psset{linewidth=0.35pt,linecolor=black}
 \pspolygon(36,0)(28,1)(22,2)(18,3)(13,5)(10,7)(7,10)(5,13)(3,18)(2,22)(1,28)(0,36)(36,36) 
 \pspolygon[linewidth=0.25pt,fillstyle=vlines,hatchangle=105,hatchwidth=0.105pt,hatchsep=44pt,hatchcolor=blue]
(13,5)(5,13)(5,36)(36,36)(36,5) 
\pspolygon[linewidth=0.25pt,fillstyle=vlines,hatchangle=15,hatchwidth=0.15pt,hatchsep=22pt,hatchcolor=green](9,9)(36,9)(36,36)(9,36)  
 \psdots[dotstyle=o](36,0)(28,1)(22,2)(18,3)(13,5)(10,7)(7,10)(5,13)(3,18)(2,22)(1,28)(0,36)
 \psdots[dotstyle=*](13,5)(5,13)
 \psdots[dotstyle=triangle*](9,9)

\rput(30,2.9){Generic} \rput(22,7){CCD} \rput(21,17){Factorial
$3^2$}

\end{pspicture}

\caption{The left graph depicts $S(D)$ and the state polyhedron for
the CCD of Example \ref{ex_designccd}. The right graph shows state
polyhedra for the three designs of Example \ref{ex_designccd}. The
empty dots correspond to vertexes/models identified by the generic
design only, while the triangle is for the sole model in the
algebraic fan of the $3^2$ design.} \label{fig_stpoly1}
\end{center}
\end{figure}

\begin{figure}[h]  
\begin{center}
\psset{unit=35.3mm,linewidth=0.4pt}
 \begin{pspicture}(-0.1,-0.1)(1.1,1.3)
     \put(0,0){}
\psset{linestyle=dashed} \psline(0,0.555555555) (0.5,1)(1,0.555555555)
\psset{linestyle=solid}
\psline(0,0) (0.0142857142857143,0.0571428571428571)
(0.0285714285714286,0.114285714285714)(0.0428571428571429,0.171428571428571)
(0.0571428571428571,0.228571428571429)(0.0714285714285714,0.285714285714286)
(0.0857142857142857,0.342857142857143) (0.1,0.4)
(0.114285714285714,0.453968253968254)(0.128571428571429,0.496825396825397)
(0.142857142857143,0.53968253968254)(0.157142857142857,0.571428571428572)
(0.171428571428571,0.603174603174603)(0.185714285714286,0.634920634920635) (0.2,0.666666666666667)
(0.214285714285714,0.690476190476191)(0.228571428571429,0.714285714285714)
(0.242857142857143,0.738095238095238)(0.257142857142857,0.761904761904762)
(0.271428571428571,0.785714285714286)(0.285714285714286,0.80952380952381) (0.3,0.822222222222222)
(0.314285714285714,0.834920634920635)(0.328571428571429,0.847619047619048)
(0.342857142857143,0.86031746031746)(0.357142857142857,0.873015873015873)
(0.371428571428571,0.885714285714286)(0.385714285714286,0.898412698412698) (0.4,0.911111111111111)
(0.414285714285714,0.915873015873016)(0.428571428571429,0.920634920634921)
(0.442857142857143,0.925396825396825)(0.457142857142857,0.93015873015873)
(0.471428571428572,0.934920634920635)(0.485714285714286,0.93968253968254) (0.5,0.944444444444444)
(0.514285714285714,0.93968253968254)(0.528571428571429,0.934920634920635)
(0.542857142857143,0.93015873015873)(0.557142857142857,0.925396825396825)
(0.571428571428571,0.920634920634921)(0.585714285714286,0.915873015873016) (0.6,0.911111111111111)
(0.614285714285714,0.898412698412699)(0.628571428571428,0.885714285714286)
(0.642857142857142,0.873015873015873)(0.657142857142857,0.860317460317461)
(0.671428571428571,0.847619047619048)(0.685714285714285,0.834920634920635)
(0.699999999999999,0.822222222222223)(0.714285714285714,0.80952380952381)
(0.728571428571428,0.785714285714287)(0.742857142857142,0.761904761904763)
(0.757142857142856,0.738095238095239)(0.771428571428571,0.714285714285716)
(0.785714285714285,0.690476190476192)(0.799999999999999,0.666666666666668)
(0.814285714285713,0.634920634920637)(0.828571428571428,0.603174603174605)
(0.842857142857142,0.571428571428574)(0.857142857142856,0.539682539682542)
(0.87142857142857,0.496825396825401)(0.885714285714284,0.453968253968258)
(0.899999999999999,0.400000000000005)(0.914285714285713,0.342857142857148)
(0.928571428571427,0.285714285714291)(0.942857142857141,0.228571428571434)
(0.957142857142856,0.171428571428577)(0.97142857142857,0.114285714285721)
(0.985714285714284,0.0571428571428636) (1,0)
\psline[linestyle=dotted](0,1)(1,1)
\psset{linewidth=0.45pt,linestyle=solid}
\rput(0.5,-0.1){$0.5$} \rput(1,-0.1){$1$} \rput(1.2,0){$w_1$}
 \psline{->}(0,0)(1.1,0)
\psline(0.5,0)(0.5,-0.05)\psline(0.25,0)(0.25,-0.05)\psline(0.75,0)(0.75,-0.05)
\psline(1,0)(1,-0.05)\psline(0,0)(0,-0.05)
\psline(0,0)(-0.05,0) \psline(0,0.25)(-0.05,0.25)
\psline(0,0.5)(-0.05,0.5) \psline(0,0.75)(-0.05,0.75)
\psline(0,1)(-0.05,1)
\psline{->}(0,-0.03125)(0,1.1)
\rput(-0.15,-0.1){$0$} \rput(-0.15,0.5){$0.5$} \rput(-0.15,1){$1$}
\rput(1.1,1.0){$3^2$} \rput(1.15,0.55){CCD}
\rput(1.15,0.25){Generic}
\end{pspicture}
\caption{Minimal aberration for three designs in two factors and
nine runs, see Example \ref{ex_designccd}.} \label{fig_abgene}
\end{center}
\end{figure}
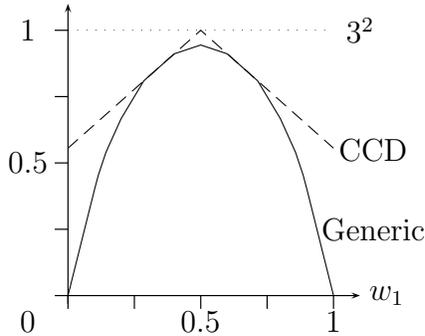

\begin{example}\label{ex_designccd} {\normalfont Consider a central composite design (CCD by Box and Wilson \cite{BW1951}) with two
factors, one observation at the origin and axial distance
$\alpha=\sqrt{2}$. The CCD has $9$ runs and its algebraic fan
contains exactly two models, namely
\begin{equation}\label{model_1}\{1,x_1,x_1^2,x_1^3,x_1^4,x_2,x_1x_2,x_1^2x_2,x_2^2\}\end{equation}
together with the model obtained by permuting the roles of $x_1$ and
$x_2$. Let $L_1$ be the set of exponents of the model support in
Equation (\ref{model_1}). Clearly, $\overline\alpha_{L_1}=(13,5)$
and the state polytope for the design ideal of the CCD is
$\operatorname{conv}\left(\{(13,5),(5,13)\}\right)$, see left graph
of Figure \ref{fig_stpoly1}. Now consider a generic design with the
same number of runs as the CCD. In \cite{CRST1999} and \cite{OS1999}
it is shown that there are $12$ corner cut models for $d=2$ and
$n=9$. By Theorem \ref{th_generic}, the algebraic fan of the generic
design contains all the $12$ corner cut models, including those in
the algebraic fan of the CCD. We consider also a full factorial
design $3^2$, which identifies only the model with support
$\{1,x_1,x_1^2\}\otimes\{1,x_2,x_2^2\}$, where $\otimes$ is the
Kronecker product. Its state polytope is the point $(9,9)$. In the
right graph of Figure \ref{fig_stpoly1} we depict the state
polyhedra for the three designs and in Figure \ref{fig_abgene} we
plot $\min_{L\in\mathcal L_a(D)}A(w,L)$ for $w=(w_1,w_2)\in[0,1]^2$
and $w_1+w_2=1$. For the CCD this is
$$\left\{\begin{array}{lcr}
\left((w_1,1-w_1)(13,5)^T\right)/9=(8w_1+5)/9&\mbox{ if }&w_1\leq 1/2\\
\left((w_1,1-w_1)(5,13)^T\right)/9=(-8w_1+13)/9&\mbox{ if }&w_1>1/2\\
\end{array}\right.$$
For the generic design the aberration curve is a piecewise linear
function with $12$ segments. Finally, the aberration for the design
$3^2$ is constant for all weights. As expected, the aberration takes
its minimum value for the generic design, over all possible weights.
 }\end{example}

\begin{figure}[h]  
\begin{center}
\psset{unit=42.3mm,linewidth=0.40pt}
 \begin{pspicture}(-0.1,-0.1)(1.1,0.9)
     \put(0,0){}
\psset{linestyle=dashed}
\psline(0,0) (0.00763358778625954,0.0229007633587786)(0.0152671755725191,0.0458015267175573)
(0.0229007633587786,0.0687022900763359)(0.0305343511450382,0.0916030534351145)
(0.0381679389312977,0.114503816793893)(0.0458015267175573,0.137404580152672)
(0.0534351145038168,0.16030534351145)(0.0610687022900763,0.183206106870229)
(0.0687022900763359,0.206106870229008)(0.0763358778625954,0.229007633587786)
(0.083969465648855,0.251908396946565)(0.0916030534351145,0.274809160305344)
(0.099236641221374,0.297709923664122)(0.106870229007634,0.320610687022901)
(0.114503816793893,0.343511450381679)(0.122137404580153,0.366412213740458)
(0.129770992366412,0.389312977099237)(0.137404580152672,0.412213740458015)
(0.145038167938931,0.432933478735006)(0.152671755725191,0.448200654307525)
(0.16030534351145,0.463467829880044)(0.16793893129771,0.478735005452563)
(0.17557251908397,0.494002181025082)(0.183206106870229,0.509269356597601)
(0.190839694656489,0.52453653217012)(0.198473282442748,0.539803707742639)
(0.206106870229008,0.550708833151581)(0.213740458015267,0.560523446019629)
(0.221374045801527,0.570338058887677)(0.229007633587786,0.580152671755725)
(0.236641221374046,0.589967284623773)(0.244274809160306,0.599781897491821)
(0.251908396946565,0.609596510359869)(0.259541984732825,0.619411123227917)
(0.267175572519084,0.629225736095965)(0.274809160305344,0.639040348964013)
(0.282442748091603,0.648854961832061)(0.290076335877863,0.658669574700109)
(0.297709923664122,0.668484187568157)(0.305343511450382,0.678298800436205)
(0.312977099236642,0.688113413304253)(0.320610687022901,0.697928026172301)
(0.328244274809161,0.707742639040349)(0.33587786259542,0.716466739367503)
(0.34351145038168,0.723009814612868)(0.351145038167939,0.729552889858234)
(0.358778625954199,0.736095965103599)(0.366412213740458,0.742639040348964)
(0.374045801526718,0.74918211559433)(0.381679389312978,0.755725190839695)
(0.389312977099237,0.76226826608506)(0.396946564885497,0.768811341330426)
(0.404580152671756,0.772082878953108)(0.412213740458016,0.773173391494002)
(0.419847328244275,0.774263904034896)(0.427480916030535,0.775354416575791)
(0.435114503816794,0.776444929116685)(0.442748091603054,0.777535441657579)
(0.450381679389314,0.778625954198473)(0.458015267175573,0.779716466739368)
(0.465648854961833,0.780806979280262)(0.473282442748092,0.781897491821156)
(0.480916030534352,0.78298800436205)(0.488549618320611,0.784078516902945)
(0.496183206106871,0.785169029443839)(0.50381679389313,0.785169029443838)
(0.51145038167939,0.784078516902944)(0.519083969465649,0.78298800436205)
(0.526717557251909,0.781897491821156)(0.534351145038169,0.780806979280262)
(0.541984732824428,0.779716466739367)(0.549618320610688,0.778625954198473)
(0.557251908396947,0.777535441657579)(0.564885496183207,0.776444929116685)
(0.572519083969466,0.775354416575791)(0.580152671755726,0.774263904034896)
(0.587786259541985,0.773173391494002)(0.595419847328245,0.772082878953108)
(0.603053435114505,0.768811341330425)(0.610687022900764,0.762268266085059)
(0.618320610687024,0.755725190839694)(0.625954198473283,0.749182115594329)
(0.633587786259543,0.742639040348963)(0.641221374045802,0.736095965103598)
(0.648854961832062,0.729552889858233)(0.656488549618321,0.723009814612867)
(0.664122137404581,0.716466739367502)(0.671755725190841,0.707742639040348)
(0.6793893129771,0.6979280261723)(0.68702290076336,0.688113413304252)(0.694656488549619,0.678298800436204)
(0.702290076335879,0.668484187568156)(0.709923664122138,0.658669574700108)
(0.717557251908398,0.64885496183206)(0.725190839694657,0.639040348964012)
(0.732824427480917,0.629225736095964)(0.740458015267177,0.619411123227916)
(0.748091603053436,0.609596510359868)(0.755725190839696,0.59978189749182)
(0.763358778625955,0.589967284623772)(0.770992366412215,0.580152671755724)
(0.778625954198474,0.570338058887676)(0.786259541984734,0.560523446019628)
(0.793893129770993,0.55070883315158)(0.801526717557253,0.539803707742637)
(0.809160305343513,0.524536532170118)(0.816793893129772,0.509269356597599)
(0.824427480916032,0.494002181025079)(0.832061068702291,0.47873500545256)
(0.839694656488551,0.463467829880041)(0.84732824427481,0.448200654307522)(0.85496183206107,0.432933478735003)
(0.862595419847329,0.412213740458012)(0.870229007633589,0.389312977099233)
(0.877862595419849,0.366412213740454)(0.885496183206108,0.343511450381676)
(0.893129770992368,0.320610687022897)(0.900763358778627,0.297709923664118)
(0.908396946564887,0.27480916030534)(0.916030534351146,0.251908396946561)
(0.923664122137406,0.229007633587782)(0.931297709923665,0.206106870229004)
(0.938931297709925,0.183206106870225)(0.946564885496185,0.160305343511446)
(0.954198473282444,0.137404580152668)(0.961832061068704,0.114503816793889)
(0.969465648854963,0.0916030534351103)(0.977099236641223,0.0687022900763317)
(0.984732824427482,0.045801526717553)(0.992366412213742,0.0229007633587743) (1,0)
\psset{linestyle=solid}
\psline(0,0) (0.00763358778625954,0.0229007633587786)(0.0152671755725191,0.0458015267175573)
(0.0229007633587786,0.0687022900763359)(0.0305343511450382,0.0916030534351145)
(0.0381679389312977,0.114503816793893)(0.0458015267175573,0.137404580152672)
(0.0534351145038168,0.16030534351145)(0.0610687022900763,0.183206106870229)
(0.0687022900763359,0.206106870229008)(0.0763358778625954,0.229007633587786)
(0.083969465648855,0.251908396946565)(0.0916030534351145,0.274809160305344)
(0.099236641221374,0.297709923664122)(0.106870229007634,0.320610687022901)
(0.114503816793893,0.343511450381679)(0.122137404580153,0.366412213740458)
(0.129770992366412,0.389312977099237)(0.137404580152672,0.412213740458015)
(0.145038167938931,0.432933478735006)(0.152671755725191,0.448200654307525)
(0.16030534351145,0.463467829880044)(0.16793893129771,0.478735005452563)
(0.17557251908397,0.494002181025082)(0.183206106870229,0.509269356597601)
(0.190839694656489,0.52453653217012)(0.198473282442748,0.539803707742639)
(0.206106870229008,0.550708833151581)(0.213740458015267,0.560523446019629)
(0.221374045801527,0.570338058887677)(0.229007633587786,0.580152671755725)
(0.236641221374046,0.589967284623773)(0.244274809160306,0.599781897491821)
(0.251908396946565,0.609596510359869)(0.259541984732825,0.619411123227917)
(0.267175572519084,0.629225736095965)(0.274809160305344,0.639040348964013)
(0.282442748091603,0.648854961832061)(0.290076335877863,0.658669574700109)
(0.297709923664122,0.668484187568157)(0.305343511450382,0.678298800436205)
(0.312977099236642,0.688113413304253)(0.320610687022901,0.697928026172301)
(0.328244274809161,0.707742639040349)(0.33587786259542,0.715376226826609)
(0.34351145038168,0.718647764449291)(0.351145038167939,0.721919302071974)
(0.358778625954199,0.725190839694657)(0.366412213740458,0.728462377317339)
(0.374045801526718,0.731733914940022)(0.381679389312978,0.735005452562705)
(0.389312977099237,0.738276990185387)(0.396946564885497,0.74154852780807)
(0.404580152671756,0.744820065430753)(0.412213740458016,0.748091603053435)
(0.419847328244275,0.751363140676118)(0.427480916030535,0.754634678298801)
(0.435114503816794,0.757906215921483)(0.442748091603054,0.761177753544166)
(0.450381679389314,0.764449291166849)(0.458015267175573,0.767720828789531)
(0.465648854961833,0.770992366412214)(0.473282442748092,0.774263904034897)
(0.480916030534352,0.777535441657579)(0.488549618320611,0.780806979280262)
(0.496183206106871,0.784078516902945)(0.50381679389313,0.784078516902944)
(0.51145038167939,0.780806979280261)(0.519083969465649,0.777535441657579)
(0.526717557251909,0.774263904034896)(0.534351145038169,0.770992366412213)
(0.541984732824428,0.767720828789531)(0.549618320610688,0.764449291166848)
(0.557251908396947,0.761177753544166)(0.564885496183207,0.757906215921483)
(0.572519083969466,0.7546346782988)(0.580152671755726,0.751363140676117)
(0.587786259541985,0.748091603053435)(0.595419847328245,0.744820065430752)
(0.603053435114505,0.741548527808069)(0.610687022900764,0.738276990185387)
(0.618320610687024,0.735005452562704)(0.625954198473283,0.731733914940021)
(0.633587786259543,0.728462377317339)(0.641221374045802,0.725190839694656)
(0.648854961832062,0.721919302071974)(0.656488549618321,0.718647764449291)
(0.664122137404581,0.715376226826608)(0.671755725190841,0.707742639040348)
(0.6793893129771,0.6979280261723)(0.68702290076336,0.688113413304252)
(0.694656488549619,0.678298800436204)(0.702290076335879,0.668484187568156)
(0.709923664122138,0.658669574700108)(0.717557251908398,0.64885496183206)
(0.725190839694657,0.639040348964012)(0.732824427480917,0.629225736095964)
(0.740458015267177,0.619411123227916)(0.748091603053436,0.609596510359868)
(0.755725190839696,0.59978189749182)(0.763358778625955,0.589967284623772)
(0.770992366412215,0.580152671755724)(0.778625954198474,0.570338058887676)
(0.786259541984734,0.560523446019628)(0.793893129770993,0.55070883315158)
(0.801526717557253,0.539803707742637)(0.809160305343513,0.524536532170118)
(0.816793893129772,0.509269356597599)(0.824427480916032,0.494002181025079)
(0.832061068702291,0.47873500545256)(0.839694656488551,0.463467829880041)
(0.84732824427481,0.448200654307522)(0.85496183206107,0.432933478735003)
(0.862595419847329,0.412213740458012)(0.870229007633589,0.389312977099233)
(0.877862595419849,0.366412213740454)(0.885496183206108,0.343511450381676)
(0.893129770992368,0.320610687022897)(0.900763358778627,0.297709923664118)
(0.908396946564887,0.27480916030534)(0.916030534351146,0.251908396946561)
(0.923664122137406,0.229007633587782)(0.931297709923665,0.206106870229004)
(0.938931297709925,0.183206106870225)(0.946564885496185,0.160305343511446)
(0.954198473282444,0.137404580152668)(0.961832061068704,0.114503816793889)
(0.969465648854963,0.0916030534351103)(0.977099236641223,0.0687022900763317)
(0.984732824427482,0.045801526717553)(0.992366412213742,0.0229007633587743) (1,0)
\psset{linewidth=0.30pt,linestyle=solid}
\rput(0.5,-0.1){$0.5$} \rput(1,-0.1){$1$} \rput(1.2,0){$w_1$}
 \psline{->}(0,0)(1.1,0)
\psline(0.5,0)(0.5,-0.05)\psline(0.25,0)(0.25,-0.05)\psline(0.75,0)(0.75,-0.05)
\psline(1,0)(1,-0.05)\psline(0,0)(0,-0.05)
\psline(0,0)(-0.05,0) \psline(0,0.2)(-0.05,0.2)
\psline(0,0.4)(-0.05,0.4) \psline(0,0.6)(-0.05,0.6)
\psline(0,0.8)(-0.05,0.8)
\psline{->}(0,-0.03125)(0,0.85)
\rput(-0.15,-0.1){$0$} \rput(-0.15,0.2){$0.2$}
\rput(-0.15,0.4){$0.4$} \rput(-0.15,0.6){$0.6$}
\rput(-0.15,0.8){$0.8$}
\end{pspicture}
\caption{Minimal aberration for $G$ (solid line) and $D$ (dashed
line), see Example \ref{ex_designonn}.} \label{fig_abonn}
\end{center}
\end{figure}
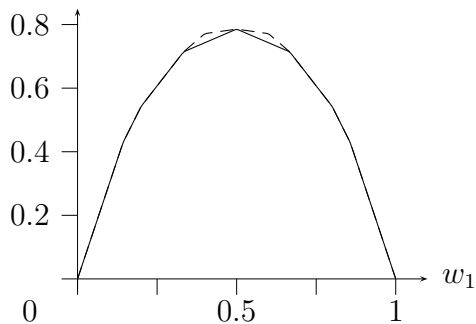

\begin{example} \label{ex_designonn}{\normalfont Consider the design $D=\{(0,0),(1,1)$, $(2,2)$, $(3,4)$,
$(5,7),(11,13)$, $(\alpha,\beta)\}$, where $(\alpha,\beta)$ $\approx
(1.82997,1.82448)$ is the only real solution of a system of
polynomial equations, see \cite[Page 47]{OS1999}. The algebraic fan
of the above design has ten models and its state polytope is
\[\operatorname{conv}\left(\{(21,0),(15,1),(11,2),(9,3),(6,5),(5,6),(3,9),(2,11),(1,15),(0,21)\}\right).\]
Now consider a generic design $G$ with the same number of runs and
factors. The algebraic fan of $G$ is the set of corner cut models
which for $7$ points in $2$ dimensions has $8$ elements, see
\cite{CRST1999} and \cite{OS1999} and thus its state polytope is the
corner cut polytope:
\[CC(7,2)=\operatorname{conv}\left(\{(21,0),(15,1),(11,2),(7,4),(4,7),(2,11),(1,15),(0,21)\}\right).\]
In Figure \ref{fig_abonn} we graph the aberration for both designs
as a function of $w_1$. Although the size of the algebraic fan of
$D$ is bigger than that for a generic design, the weighted
aberration takes minimal value for the generic design for all
possible weight vectors $(w_1,1-w_1)$.}\end{example}

\begin{example} {\normalfont The aberration of some sets of
multi-indices does not depend on $w$. For instance, consider the
following sets in two dimensions
$$\begin{array}{ccc}L_n&=&\{(i,i):i=0,\ldots,n-1\}\\
M_n&=&\{(i,j):i,j=0,\ldots,n-1\}\\
N_n&=&\{(i,j):0\leq i+j\leq n\}
\end{array}$$ for which the aberrations are $A(w,L_n)=(n-1)/2$,
$A(w,M_n)=(n-1)/2$ and $A(w,N_n)=n/3$. To properly compare the above
aberrations, the sets $L,M,N$ must have the same size. Below are
values $m,n$ and $p=m^2$ such that $\#L_p=\#M_m=\#N_n$ for $m$ up to
$8000$.
$$\begin{array}{rrrrr}m&n&A(w,L_p)&A(w,M_m)&A(w,N_n)\\ \hline
1&0&0&0&0\\
6&8&17.5&2.5&2.6\\
35&49&612.0&17.0&16.3\\
204&288&20807.5&101.5&96.0\\
1189&1981&7.0\times 10^5&594.0&560.3\\
6930&9800&2.4\times 10^7&3464.5&3266.6\\
40391&57121&8.1\times 10^8&20195.0&19040.3\\
\end{array}$$
As sample size grows, the aberration of the triangular set $N_n$
remains smaller than for the square set $M_m$. }\end{example}

\subsection{Bounds for the aberration}
Although the minimal value of the aberration $A^*(w,n)$, depends on
the weight vector $w=(w_1,\ldots,w_d)$, we can carry out a special
normalisation which leads to bounds for the minimal aberration.
These bounds depend only on a simple function of the weights,
surprisingly the geometric mean. Our construction is based upon the
expected value of auxiliary random variables which are suitably
constructed.


For the rest of this Section let $D\subset \mathbb R^d$ be a generic
design with $n$ points.
Let $w$ be a fixed weight vector with positive elements and let $L$
be the corner cut model identified by $w$. We recall that $|L|=n$.

\begin{figure}[h]  
\begin{center}
\psset{unit=8.3mm,linewidth=0.8pt}
 \begin{pspicture}(-2,-2)(4,4.5)
     \put(0,0){}
\psset{linecolor=black,linewidth=0.25pt}
     \psline{->}(0,-0.5)(0,4.0)
     \psline{->}(-0.5,0)(3.0,0)
     \pspolygon[linewidth=0.25pt,fillstyle=vlines,hatchangle=65,hatchwidth=0.25pt,hatchsep=4pt,hatchcolor=darkgray](1,0)(2,0)(2,1)(1,1)
     \pspolygon[linewidth=0.25pt,fillstyle=vlines,hatchangle=65,hatchwidth=0.25pt,hatchsep=4pt,hatchcolor=darkgray](1,1)(2,1)(2,2)(1,2)
     \pspolygon[linewidth=0.25pt,fillstyle=vlines,hatchangle=65,hatchwidth=0.25pt,hatchsep=4pt,hatchcolor=darkgray](0,2)(1,2)(1,3)(0,3)
     \pspolygon[linewidth=0.25pt,fillstyle=vlines,hatchangle=65,hatchwidth=0.25pt,hatchsep=4pt,hatchcolor=darkgray](0,1)(1,1)(1,2)(0,2)
     \pspolygon[linewidth=0.25pt,fillstyle=vlines,hatchangle=65,hatchwidth=0.25pt,hatchsep=4pt,hatchcolor=darkgray](0,0)(1,0)(1,1)(0,1)

     \pspolygon[linecolor=blue,linewidth=0.5pt,fillstyle=vlines,hatchangle=158,hatchwidth=0.5pt,hatchsep=12pt,hatchcolor=blue](0,0)(2.7116,0)(0,3.6878)
     \rput(1,-0.5){\tiny{$1$}} \rput(2,-0.5){\tiny{$2$}}
     \rput(-0.5,1){\tiny{$1$}} \rput(-0.5,2){\tiny{$2$}} \rput(-0.5,3){\tiny{$3$}}
     \rput(-0.35,-0.5){\tiny{$0$}}
     \psset{linewidth=0.6pt}
\psline(-0.4,3.1)(2.1,-0.3)
\psline[linestyle=dotted,linewidth=1.3pt,linecolor=darkgray]{->}(0,0)(2.55,1.875)\rput(2.8,2.1){$w$}
\rput(2.8,0.75){${S}(w)$} \rput(3.3,0){$v_1$} \rput(0,4.2){$v_2$}
     \pscircle*[fillcolor=black](0,0){0.1}
     \pscircle*[fillcolor=black](1,0){0.1}
     \pscircle*[fillcolor=black](0,1){0.1}
     \pscircle*[fillcolor=black](1,1){0.1}
     \pscircle*[fillcolor=black](0,2){0.1}
\end{pspicture}
 \begin{pspicture}(-3,-2)(3,4.5)
     \put(0,0){}
\psset{linecolor=black,linewidth=0.25pt}
     \psline{->}(0,-0.5)(0,4.0)
     \psline{->}(-0.5,0)(3.0,0)

     \pspolygon[linewidth=0.25pt,fillstyle=vlines,hatchangle=65,hatchwidth=0.25pt,hatchsep=4pt,hatchcolor=darkgray](0,0)(-1,0)(-1,-1)(0,-1)
     \pspolygon[linewidth=0.25pt,fillstyle=vlines,hatchangle=65,hatchwidth=0.25pt,hatchsep=4pt,hatchcolor=darkgray](1,0)(0,0)(0,-1)(1,-1)
     \pspolygon[linewidth=0.25pt,fillstyle=vlines,hatchangle=65,hatchwidth=0.25pt,hatchsep=4pt,hatchcolor=darkgray](0,0)(1,0)(1,1)(0,1)
     \pspolygon[linewidth=0.25pt,fillstyle=vlines,hatchangle=65,hatchwidth=0.25pt,hatchsep=4pt,hatchcolor=darkgray](0,1)(-1,1)(-1,0)(0,0)
     \pspolygon[linewidth=0.25pt,fillstyle=vlines,hatchangle=65,hatchwidth=0.25pt,hatchsep=4pt,hatchcolor=darkgray](0,2)(-1,2)(-1,1)(0,1)

     \pspolygon[linecolor=blue,linewidth=0.5pt,fillstyle=vlines,hatchangle=158,hatchwidth=0.5pt,hatchsep=12pt,hatchcolor=blue](-1,-1)(1.7116,-1)(-1,2.6878)

     \rput(1,-0.5){\tiny{$1$}} \rput(2,-0.5){\tiny{$2$}}
     \rput(-0.5,1){\tiny{$1$}} \rput(-0.5,2){\tiny{$2$}} \rput(-0.5,3){\tiny{$3$}}
     \rput(-0.35,-0.5){\tiny{$0$}}
     \psset{linewidth=0.6pt}
\psline(-0.4,3.1)(2.1,-0.3)
\psline[linestyle=dotted,linewidth=1.3pt,linecolor=darkgray]{->}(0,0)(2.55,1.875)\rput(2.8,2.1){$w$}
\rput(2,-1.3){$\underline{S}(w)$} \rput(3.3,0){$v_1$}
\rput(0,4.2){$v_2$}

     \pscircle*[fillcolor=black](0,0){0.1}
     \pscircle*[fillcolor=black](1,0){0.1}
     \pscircle*[fillcolor=black](0,1){0.1}
     \pscircle*[fillcolor=black](1,1){0.1}
     \pscircle*[fillcolor=black](0,2){0.1}

\end{pspicture}

\caption{Bidimensional corner cut together with upper (left diagram)
and lower cells (right diagram) $\overline Q$ and $\underline Q$. In
both diagrams the vector $w$, a separating hyperplane and equivalent
simplexes $S(w)$ and $\underline{S}(w)$ were added.} \label{fig_cc2}

\end{center}
\end{figure}
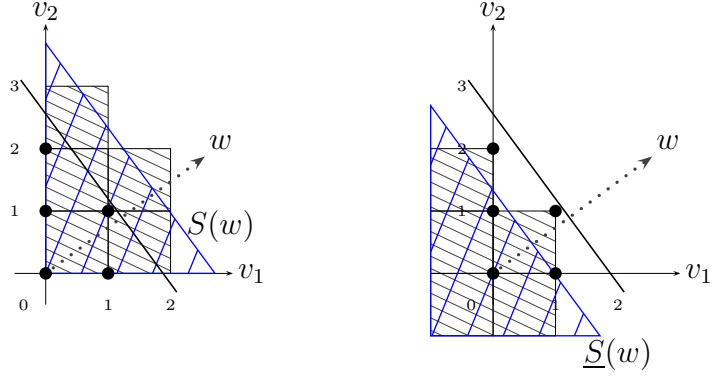

For an integer multindex $\alpha$ define its {\em upper cell} as the
unit cube with lower vertex at $\alpha$
$$\overline{c}(\alpha) = \{v\in\mathbb R^d: \alpha_i\leq v_i\leq \alpha_i+1\}$$
and similarly the {\em lower cell} of $\alpha$ is
$$\underline{c}(\alpha) = \{v\in\mathbb R^d: \alpha_i-1\leq v_i\leq \alpha_i\}$$
Define:
$$\underline{Q} = \cup_{\alpha \in L}\; \underline{c}(\alpha), \; \; \overline{Q} = \cup_{\alpha \in L} \; \overline{c}(\alpha).$$
See Figure \ref{fig_cc2} for a depiction of lower and upper cells
with $L$ a corner cut.


Clearly, the volume of $\overline{Q}$ and of $\underline{Q}$ equals
$n$, that is the cardinality of $L$. We now create a simplex
$S(w)\subset\mathbb R^d$ which is directed by the vector $w$ and has
volume $n$. We call this simplex and the subset of the first orthant
below it the {\em equivalent simplex}, which is formally
$S(w)=\left\{v\in\mathbb R^d_{\geq 0}: \sum_{i=1}^d{v_iw_i}\leq c
\right \}$. The volume of $S(w)$ is determined up to the constant
$c>0$. We find the value of this constant by setting the total
volume of the equivalent simplex equal to $n$:
$$n=\frac{c^d}{d! \prod_{i=1}^d w_i},$$ giving
\begin{equation}\label{eq_c}c= \left( n d! \right)^{\frac{1}{d}}g(w),\end{equation} where
$$g(w) = \left(\prod_{i=1}^d w_i\right)^{\frac{1}{d}}$$
is the geometric mean of the components of the weight vector $w$. We
call $H(w)$ the hyperplane which limits the equivalent simplex, that
is $H(w)=\left\{v\in\mathbb R^d_{\geq 0}: \sum_{i=1}^d{v_iw_i}= c
\right \}$.

The expected value of a random variable with uniform support over
$S(w)$ will be used now to compute bounds for aberration. We can
compute a notional value of $A$, the linear aberration for a
distribution $D$ as the expectation  $A(w,S(w))=\mbox{E}(\sum w_i
X_i)$ for the random vector $(X_1,\ldots ,X_d)$ with uniform
distribution over $S(w)$. Thus for the equivalent simplex we have
that
\begin{equation}\label{eq_A+}
A(w,S(w))=\frac{1}{n} \frac{d}{(d+1)!}\frac{c^{d+1}}{\prod_{i=1}^d
w_i} = (nd!)^{\frac{1}{d}} \frac{d}{d+1}  g(w),
\end{equation}
after substituting Equation (\ref{eq_c}) in $A(w,S(w))$

We observe that the region $\underline{Q}$ is obtained from
$\overline{Q}$ by a negative shift $(-1,\ldots, -1)$. As before, we
consider a random vector with joint uniform distribution over
$\underline{Q}$. We then use the expected value of $\sum w_i X_i$ as
the aberration $A(w,\underline{Q})$. Analogously we define
$A(w,\overline Q)$ and we have
$$A(w,\underline{Q}) = A(w,\overline{Q}) -1$$
Similarly we can create a region $\underline{S}(w)$ by the same
downward shift, and we have
$$A(w,\underline{S}(w)) = A(w,{S}(w)) -1.$$
As $D$ is generic and thus $L$ is a corner cut there exist cutting
hyperplanes separating $L$ from its complement in $\mathbb Z_{\geq
0}^d$. Moreover if $w$ is in the interior of the normal cone of the
corner cut polytope, then we can select a cutting hyperplane $H$
which is orthogonal to $w$ and thus parallel to $H(w)$, see
\cite{OS1999}.

\begin{example}{\normalfont
Consider a generic design with $d=2,n=3$ and $L=\{(0,0)$,
$(1,0)$,$(2,0)\}$. The weight vector $w=(1,2)$ is not in the
interior of a normal cone of the corner cut polytope $CC(2,3)$.
Indeed the weight vector is on the boundary of the normal cone
separating $L$ from the corner cut model $\{(0,0),(1,0),(0,1)\}$.
The hyperplanes perpendicular to $w$ are $2x_1-x_2=c$ and none of
them is a cutting hyperplane for $L$. }\end{example}

 By a simple argument the simplex
$S_H$ with faces $x_i=0,\; (i=1,\ldots,d)$ and $H$ lies wholly
within the upper quadrant region $\overline{Q}$ because otherwise,
the cutting hyperplane hypothesis for $H$ would be violated and thus
$S_H$ has volume less than $n$. Recall that the equivalent simplex
$S(w)$ has volume $n$.

There is one additional argument that leads to our first inequality.
Since the region $\overline{Q}$ and the equivalent simplex $S(w)$
have the same volume $n$, it must be that $\overline{Q}$ protrudes
beyond $S(w)$. Equivalently we may move mass from $\overline Q$,
that is, beyond $H(w)$, inside $S(w)$. As this mass occurs
orthogonally to $w$, we claim that this movement diminishes the
aberration, thus
$$A(w,S(w))\leq A(w,\overline Q).$$ This property is also inherited
by the downward shifted version, and we have $A(w,\underline
S(w))\leq A(w,\underline Q).$ The same orthogonality argument shows
the middle inequality in the following sequence:
$$A(w,\underline{S}(w))  \leq A(w,\underline{Q}) \leq A(w,S(w)) \leq A(w,\overline{Q}).$$
By Theorem \ref{th_exist}, as the design is generic and $L$ is the
model identified by $w$, clearly we have $$A(w,\underline Q)\leq
A^*(w,n)\leq A(w,\overline Q).$$ Analogous argument and construction
as above shows that $A(w,\overline Q)\leq A(w,S(w))+1$.

\begin{theorem} \label{theo_bounds}
Let $D\subset\mathbb R^d $ be a generic design with $n$ points; let
$w\in\mathbb R^d$ be a vector of positive weights. Then the minimal
aberration $A^*(w,n)$ satisfies the bounds
\begin{equation}\label{ec_bounds} A(w,S(w))-1 \leq A^*(w,n) \leq A(w,S(w))
+1,\end{equation} where $A(w,S(w))$ is computed in Equation
(\ref{eq_A+}).
\end{theorem}

There are various kinds of asymptotic that this formula leads to.
From the inequality between geometric and arithmetic mean we have
$g(w) \leq \frac{1}{d}$. This suggests the condition
$\lim_{d\mapsto\infty} g(w)=c/d$
for some constant $0 \leq c \leq 1$. Now for $w_i = (1+\delta_i)/d$,
with $\sum \delta_i = 0$, and assuming convergence of $\sum
\delta_i^2$ and $n = k^d$, we use use Stirling's approximation to
obtain
$$\lim_{d\mapsto\infty}A^*(w,n)=\frac{kc}{e}.$$
Such limits may be considered as asymptotic identifiability rates,
analogous to the more familiar Nyquist rates in Fourier analysis.

\begin{figure}[h]
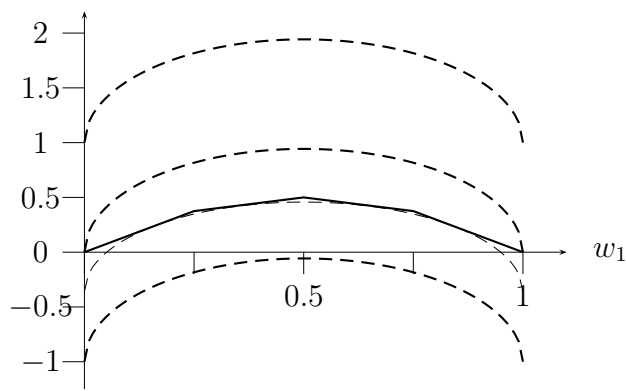
  
\begin{center}
\psset{unit=58.3mm,linewidth=0.8pt}


\caption{Minimal aberration $A^*(w,n)$ (solid line) for a generic
design with $d=2$, $n=4$; bounds $A(w,S(w))$ and $A(w,S(w))\pm1$ of
Theorem \ref{theo_bounds} (dashed lines). We also show approximate
aberration $\tilde A$ using Theorem \ref{th_apr_ab} (thin dashed
line).} \label{fig_ab2}

\end{center}
\end{figure}

\begin{figure}[h]
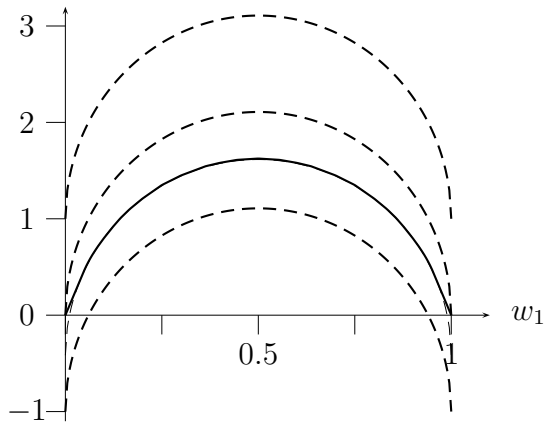
  
\begin{center}
\psset{unit=51.3mm,linewidth=0.8pt}
 %
\caption{Minimal aberration $A^*(w,n)$ (solid line) for a generic
design with $d=2$, $n=20$; bounds $A(w,S(w))$ and $A(w,S(w))\pm 1$
and (dashed lines) of Theorem \ref{theo_bounds}. The figure also
shows approximate aberration $\tilde A$ of Theorem \ref{th_apr_ab}
(thin dashed line) which almost overlaps the solid line.}
\label{fig_ab3}

\end{center}
\end{figure}

\begin{figure}[h]
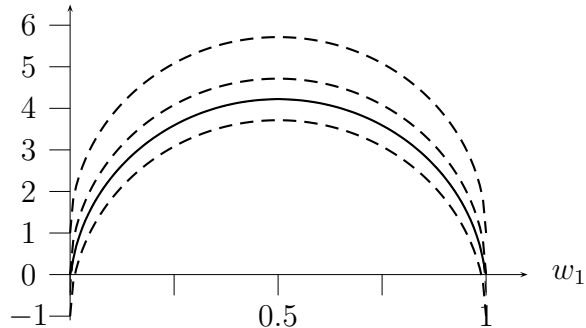
  
\begin{center}
\psset{unit=55.3mm,linewidth=0.8pt}
 %
\caption{Minimal aberration $A^*(w,n)$ (solid line) for a generic
design with $d=2$, $n=100$; bounds $A(w,S(w))$ and $A(w,S(s))\pm1$
(dashed lines). The approximate aberration $\tilde A$ of Equation
(\ref{eq_minab}) (thin dashed line) is also plotted, but is
undistinguishable from the minimal aberration.} \label{fig_ab4}
\end{center}
\end{figure}

\begin{example}{\normalfont
For small $d$ and $n$ the bounds of Equation (\ref{ec_bounds}) are
rather coarse. Figure \ref{fig_ab2} shows the bounds $A(w,S(w))\pm1$
of Theorem \ref{theo_bounds} together with the minimal aberration
$A^*(w,n)$, plotted as function of $w_1$ for $d=2$ and $n=4$. Notice
that, as function of $w$, the minimal aberration $A^*(w,n)$ is a
piece-wise linear graph (this is a general fact, consequence of
Definition \ref{def_ab}), each segment corresponding to a different
vertex (different corner cut) of the corner cut polytope. Figures
\ref{fig_ab3} and \ref{fig_ab4} give the bounds and minimal
aberration for $n=20$ and $n=100$. In Figures \ref{fig_ab2},
\ref{fig_ab3} and \ref{fig_ab4} we also added a curve for the
approximate aberration which is presented in Theorem \ref{th_apr_ab}
below. }\end{example}


\subsection{Approximated state polytope for generic designs}
\label{sec_ap_sp}

Note that as $w$ changes the hyperplanes $H(w)$ are tangent to the
surface defined by
$$ \prod_{i=1}^d x_i = c^d = n d! \left(\frac{1}{d}\right)^d$$
and the (normalised) centroids of the equivalent simplices lie on
the surface defined by
\begin{equation} \label{ec_centroid_surf}
\prod_{i=1}^d x_i = b^+ = n \left(\frac{1}{d+1}\right)^d d!
\end{equation}

We can solve an equivalent optimisation problem to the computations
of $A(w,S(w))$ in terms of the tangent surfaces: for all centroids
lying above or on the surface of Equation (\ref{ec_centroid_surf}),
the minimum value of $A(w,S(w))$ is achieved at the centroid of the
tangent.

In the above argument, we are essentially using the surface in
Equation (\ref{ec_centroid_surf}) to approximate the lower border of
the state polytope for a generic design, i.e. the lower border of
the corner cut polytope. In order to improve the bounds given in
Theorem \ref{theo_bounds}, it seems natural simply to take a surface
defined by
\begin{equation}\label{ec_aprox}\prod_{i=1}^d(x_i+a) = b\end{equation}
with fixed $a,b$. In Theorem \ref{theo_bounds}, we have $a = \pm 1$
and $b = b^+$ in Equation (\ref{ec_centroid_surf}). In Appendix B we
discuss an approach to select the values $a,b$ to obtain a good
approximation of the corner cut polytope.

The following theorem estimates minimal aberration for generic
designs using the approximation of Equation (\ref{ec_aprox}). The
proof is based on simple ideas of constrained optimization, see
Appendix A.

\begin{theorem} \label{th_apr_ab}
Let $w=(w_1,\ldots,w_d)$ be a fixed positive weight vector; let
$D\subset\mathbb R^d$ be a generic design with $n$ points. Let the
state polytope of $I(D)$ be approximated by Equation
(\ref{ec_aprox}). Then the value
\begin{equation}\label{eq_minab}\tilde A(w)=db^{1/d}g(w)-a\sum_{i=1}^d{w_i}\end{equation}
is an approximation of $A^*(w,n)$.
\end{theorem}
We recall that $g(w)$ is the geometrical mean of the components in
$w$. Figures \ref{fig_ab2}, \ref{fig_ab3} and \ref{fig_ab4} give
examples ($d=2$ factors, $n=4,20,100$) of the minimal aberration
$\tilde A(w)$ in Theorem \ref{th_apr_ab}. The values $a,b$ for each
case were selected using the technique in Appendix B.

\section{Examples}\label{sec_Ex}

In this section we discuss through extended examples other possible
uses of the ideas on generic designs and aberration. In Section
\ref{sec_LHS} we explore and conjecture the existence of generic
designs over Latin hypercubes for all factors and sample sizes. In
Section \ref{sec_2kp} we compare fractional factorial designs
through their state polytopes.

\subsection{Latin hypercube design}\label{sec_LHS}

Latin hypercube designs (LH) were first proposed by McKay \textit{et
al}. \cite{MBC1979} in the context of computer experiments. Latin
hypercubes are designs with reasonable space filling properties and
good projections in lower dimensions.

Theorem \ref{th_exist} relates minimal aberration to generic
designs, i.e. if the design is generic, then it identifies models of
lower weighted degree (and minimal aberration) for any weight vector
$w$. In what follows we study LH using Definition \ref{def_generic}
of generic designs.

The construction of a Latin hypercube design can be summarised as
follows.\begin{enumerate}
\item Divide the range of each factor into $n$ equal segments.
\item\label{step_2} Select a value in each segment using a random
uniform distribution, or any other continuous distribution.
\item Randomly permute the list for each factor.
\end{enumerate}

By Theorem 30 in \cite{PRW2000}, a Latin hypercube design
constructed as above is generic with probability one.

We now consider a special type of LH designs. This type is
constructed by selecting a fixed value in every segment in Step
\ref{step_2}. For instance, we could select the minimum, maximum or
the midpoint value for every segment.
\begin{figure}
\begin{center}
\psset{unit=4.5mm}
 \begin{pspicture}(-1,-1)(15,13)
\put(0,0){}

 \psset{linecolor=gray,linewidth=0.95pt} \psline{->}(2,0)(16,0) \psline{->}(2,0)(2,11)
 \psset{linewidth=0.25pt}

\psline(2,10)(16,10) \psline(2,9)(16,9) \psline(2,8)(16,8)
\psline(2,7)(16,7) \psline(2,6)(16,6) \psline(2,5)(16,5)
\psline(2,4)(16,4) \psline(2,3)(16,3) \psline(2,2)(16,2)
\psline(2,1)(16,1)

\psline(3,0)(3,10) \psline(4,0)(4,10) \psline(4,0)(4,10)
\psline(5,0)(5,10) \psline(6,0)(6,10) \psline(7,0)(7,10)
\psline(8,0)(8,10) \psline(9,0)(9,10) \psline(10,0)(10,10)
\psline(11,0)(11,10) \psline(12,0)(12,10) \psline(13,0)(13,10)
\psline(14,0)(14,10) \psline(15,0)(15,10)

 \psset{linecolor=black,linewidth=1pt}

 \psline (2,10) (3,3.33333333333333)
(4,8.33333333333333) (5,5.33333333333333) (6,6.44444444444444)
(7,7.77777777777778) (8,9.51240079365079) (9,9.51413333333333)
(10,9.84310437109724) (11,9.99046263345196) (12,9.99096551724138)
(13,9.98) (14,9.98) (15,9.99844224316984) \psline[linestyle=dashed]
(2,10) (3,3.33333333333333) (4,6.66666666666667)
(5,1.33333333333333) (6,6.22222222222222) (7,5.15873015873016)
(8,8.73263888888889) (9,9.17693333333333) (10,9.84228367528992)
(11,9.97245551601424) (12,9.9891724137931) (13,9.98) (14,9.98)
(15,9.99844224316984)

\put(0.1,9.8){\small{$100\%$}} \put(0.1,4.8){\small{$75\%$}}
\put(0.1,-0.2){\small{$50\%$}} \put(0.1,8.8){\small{$95\%$}}
\put(16.4,-0.1){\small{$n$}} \put(1.9,-0.9){\small{$2$}}
\put(2.9,-0.9){\small{$3$}} \put(3.9,-0.9){\small{$4$}}
\put(4.9,-0.9){\small{$5$}} \put(5.9,-0.9){\small{$6$}}
\put(6.9,-0.9){\small{$7$}} \put(7.9,-0.9){\small{$8$}}
\put(8.9,-0.9){\small{$9$}} \put(9.7,-0.9){\small{$10$}}
\put(10.7,-0.9){\small{$11$}} \put(11.7,-0.9){\small{$12$}}
 \put(12.7,-0.9){\small{$13$}} \put(13.7,-0.9){\small{$14$}} \put(14.7,-0.9){\small{$15$}}
\put(5,9.2){\small{Generic}} \psline{->}(6.5,9.2)(7.6,9)
\put(8.6,5.2){\small{Maximal fan}} \psline{->}(8.5,5.3)(7.4,5.9)
\end{pspicture}
\caption{ Percentage of generic LHS designs for $d=2$ and $n\leq
15$. }\label{fig_LHS}
\end{center}

\end{figure}
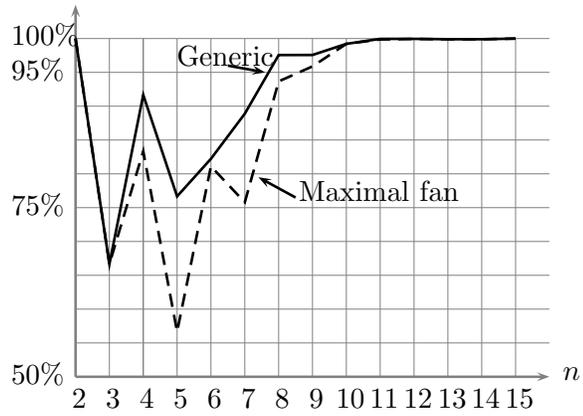
\begin{figure}
\begin{center}

\psset{unit=5.5mm,linewidth=0.7pt}
 \begin{pspicture}(3,-1)(15,5)
\put(0,0){}

%

\psline (3,0.477121254719662) (4,1.07918124604762)(5,0.632023214705406) (6,0.7501225267834) (7,0.954242509439325)
(8,1.61296700560533) (9,1.61451289075472) (10,2.10541915131129)(11,3.32160152154028) (12,3.34512670224318) (13,3.36921585741014)
(14,3.59423465381592) (15,4.10853033151892)

\psline[linestyle=dashed] (3,0.477121254719662)(4,0.778151250383644) (5,0.363177902412826) (6,0.723793588061051)
(7,0.616070705770777) (8,1.19812961896674) (9,1.38559498208857)(10,2.1031533475819) (11,2.86099535488618) (12,3.26649834548973)
(13,3.36921585741014) (14,3.59423465381592) (15,4.10853033151892)
\psline(3,0)(3,5)
\psline{->}(3,0)(15.5,0)
\psset{linestyle=dashed,linewidth=0.3pt}
 \psline(3,1)(15,1) \psline(3,2)(15,2)
\psline(3,3)(15,3) \psline(3,4)(15,4) \psline(3,5)(15,5)

\psset{linestyle=solid} \psline(4,0)(4,-0.1) \psline(5,0)(5,-0.1)
\psline(6,0)(6,-0.1) \psline(7,0)(7,-0.1) \psline(8,0)(8,-0.1)
\psline(9,0)(9,-0.1) \psline(10,0)(10,-0.1) \psline(11,0)(11,-0.1)
\psline(12,0)(12,-0.1) \psline(13,0)(13,-0.1) \psline(14,0)(14,-0.1)
\psline(15,0)(15,-0.1)

\rput(3,-0.4){$3$} \rput(5,-0.4){$5$} \rput(9.9,-0.4){$10$}
\rput(14.9,-0.4){$15$} \rput(15.9,0){$n$}

\rput(2,1){\tiny{$10^{-1}$}}
\rput(2,2){\tiny{$10^{-2}$}}\rput(2,3){\tiny{$10^{-3}$}}
\rput(2,4){\tiny{$10^{-4}$}} \rput(2,5){\tiny{$10^{-5}$}}

\psline{->}(6.6,2.5)(7.5,1.5) \psline{->}(10,0.5)(8.7,1.2)

 \rput(5,2.5){Generic}
 \rput(12,0.5){Maximal fan}

\end{pspicture}
\caption{ Minus logarithm of the percentage of non generic LHS
designs for $d=2$ and $n\leq 15$. }\label{fig_LHS2}
\end{center}
\end{figure}
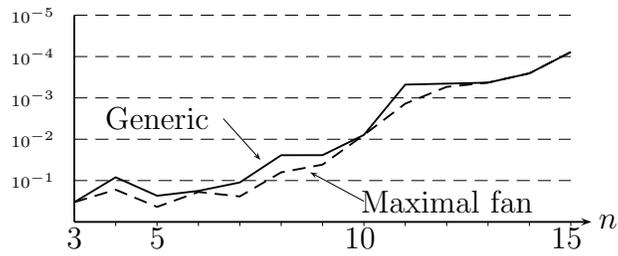

There are a few obvious cases of LH designs which are not generic,
for example when the points of the design lie on a line. We have
performed exhaustive search for a few cases of LH in two dimensions.
 Our search
points out to the existence of generic LH for different values of
$d,n$. In fact for the values we tried the proportion of generic LH
tends clearly to one.
 See Figures
\ref{fig_LHS} and \ref{fig_LHS2} for a depiction of the results,
where we additionally plot the proportion of {\em maximal fan}
designs among LH, i.e. LH designs that identify all possible
staircase models for given $d,n$. We have the following conjecture
for the existence of generic LHS for any value of $d,n$.

\begin{conjecture}
For every $d\geq 2$ and $n\geq 2$ there exists at least one generic
LH design, constructed by setting a fixed value for every one of the
$n$ segments in the above procedure.
\end{conjecture}

\begin{figure}
\begin{center}

\psset{unit=2.7mm,linewidth=0.5pt}
 \begin{pspicture}(-2.5,-1)(10,10)   
\psset{linecolor=gray}
 \psline{->}(0,0)(10,0)
 \psline{->}(0,0)(0,10)
 \psline(0,0)(9,0)(9,9)(0,9)(0,0)
\psset{linecolor=black}
  \pscircle*[fillcolor=black](0,7){0.15}
 \pscircle*[fillcolor=black](1,8){0.15}
 \pscircle*[fillcolor=black](2,9){0.15}
 \pscircle*[fillcolor=black](3,0){0.15}
 \pscircle*[fillcolor=black](4,1){0.15}
 \pscircle*[fillcolor=black](5,4){0.15}
 \pscircle*[fillcolor=black](6,2){0.15}
 \pscircle*[fillcolor=black](7,3){0.15}
 \pscircle*[fillcolor=black](8,5){0.15}
 \pscircle*[fillcolor=black](9,6){0.15}

\end{pspicture}
 \begin{pspicture}(-2.5,-1)(10,10)   
\psset{linecolor=gray}
 \psline{->}(0,0)(10,0)
 \psline{->}(0,0)(0,10)
 \psline(0,0)(9,0)(9,9)(0,9)(0,0)
\psset{linecolor=black}

  \pscircle*[fillcolor=black](0,8){0.15}
 \pscircle*[fillcolor=black](1,9){0.15}
 \pscircle*[fillcolor=black](2,0){0.15}
 \pscircle*[fillcolor=black](3,6){0.15}
 \pscircle*[fillcolor=black](4,1){0.15}
 \pscircle*[fillcolor=black](5,2){0.15}
 \pscircle*[fillcolor=black](6,3){0.15}
 \pscircle*[fillcolor=black](7,4){0.15}
 \pscircle*[fillcolor=black](8,5){0.15}
 \pscircle*[fillcolor=black](9,7){0.15}
\end{pspicture}
 \begin{pspicture}(-2.5,-1)(10,10)   
\psset{linecolor=gray}
 \psline{->}(0,0)(10,0)
 \psline{->}(0,0)(0,10)
 \psline(0,0)(9,0)(9,9)(0,9)(0,0)
\psset{linecolor=black}
  \pscircle*[fillcolor=black](0,8){0.15}
 \pscircle*[fillcolor=black](1,9){0.15}
 \pscircle*[fillcolor=black](2,3){0.15}
 \pscircle*[fillcolor=black](3,4){0.15}
 \pscircle*[fillcolor=black](4,0){0.15}
 \pscircle*[fillcolor=black](5,6){0.15}
 \pscircle*[fillcolor=black](6,7){0.15}
 \pscircle*[fillcolor=black](7,1){0.15}
 \pscircle*[fillcolor=black](8,2){0.15}
 \pscircle*[fillcolor=black](9,5){0.15}

\end{pspicture}
\begin{pspicture}(-2.5,-1)(10,10)   
\psset{linecolor=gray}
 \psline{->}(0,0)(10,0)
 \psline{->}(0,0)(0,10)
 \psline(0,0)(9,0)(9,9)(0,9)(0,0)
\psset{linecolor=black}
  \pscircle*[fillcolor=black](0,3){0.15}
 \pscircle*[fillcolor=black](1,4){0.15}
 \pscircle*[fillcolor=black](2,2){0.15}
 \pscircle*[fillcolor=black](3,6){0.15}
 \pscircle*[fillcolor=black](4,1){0.15}
 \pscircle*[fillcolor=black](5,5){0.15}
 \pscircle*[fillcolor=black](6,9){0.15}
 \pscircle*[fillcolor=black](7,7){0.15}
 \pscircle*[fillcolor=black](8,8){0.15}
 \pscircle*[fillcolor=black](9,0){0.15}

\end{pspicture}
\begin{pspicture}(-2.5,-1)(10,10)   

\psset{linecolor=gray}
 \psline{->}(0,0)(10,0)
 \psline{->}(0,0)(0,10)
 \psline(0,0)(9,0)(9,9)(0,9)(0,0)
\psset{linecolor=black}

  \pscircle*[fillcolor=black](0,9){0.15}
 \pscircle*[fillcolor=black](1,6){0.15}
 \pscircle*[fillcolor=black](2,7){0.15}
 \pscircle*[fillcolor=black](3,8){0.15}
 \pscircle*[fillcolor=black](4,5){0.15}
 \pscircle*[fillcolor=black](5,2){0.15}
 \pscircle*[fillcolor=black](6,1){0.15}
 \pscircle*[fillcolor=black](7,4){0.15}
 \pscircle*[fillcolor=black](8,3){0.15}
 \pscircle*[fillcolor=black](9,0){0.15}

\end{pspicture}
\begin{pspicture}(-2.5,-1)(10,10)   

\psset{linecolor=gray}
 \psline{->}(0,0)(10,0)
 \psline{->}(0,0)(0,10)
 \psline(0,0)(9,0)(9,9)(0,9)(0,0)
\psset{linecolor=black}

  \pscircle*[fillcolor=black](0,1){0.15}
 \pscircle*[fillcolor=black](1,9){0.15}
 \pscircle*[fillcolor=black](2,7){0.15}
 \pscircle*[fillcolor=black](3,6){0.15}
 \pscircle*[fillcolor=black](4,5){0.15}
 \pscircle*[fillcolor=black](5,8){0.15}
 \pscircle*[fillcolor=black](6,4){0.15}
 \pscircle*[fillcolor=black](7,3){0.15}
 \pscircle*[fillcolor=black](8,2){0.15}
 \pscircle*[fillcolor=black](9,0){0.15}

\end{pspicture}

\caption{LH on $[0,1]^2$ for $d=2,n=10$ which are not generic and
identify $L$.}\label{fig_nongenLHS}
\end{center}
\end{figure}
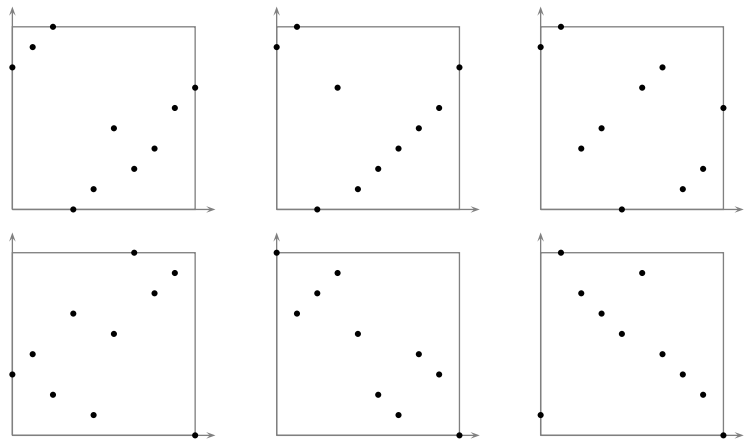

Experimentally we observed that when the sample size is
$n={k+1\choose d}$ for $k\geq 1$, the genericity of a LH design is
closely linked to the identification of a model of total degree
$k-1$. For example for $k=4,d=2,n=10$ there are $10!$ LH of which
$99\%$ are generic. Of the remaining $1\%$ which are not generic
only $6$ designs (up to reflection and rotation), which are given in
Figure \ref{fig_nongenLHS}, identify the cubic model with exponent
set
$$L=\{(0,0),(1,0),(0,1),(2,0),(1,1),(0,2),(3,0),(2,1),(1,2),(0,3)\}.$$

\subsection{Orthogonal fractions}\label{sec_2kp}

In this Section we consider some of the techniques of this paper for
the class of fractional factorial designs with two levels. We first
explore the relation between state polyhedron and then later propose
a tool to compare the identification capability of designs.

In Examples \ref{ex_designccd} and \ref{ex_designonn} we observed
that in general, nesting of state polyhedra for two designs does not
imply any easy relation between the algebraic fan of the designs. If
instead we restrict to the family of designs with two levels then
there is a clear relation between such nesting and algebraic fans.
We have the following Lemma from Chapter 6 in the Ph.D. thesis by
Maruri-Aguilar (2007).

\begin{lemma} \label{lem_2k}
Let $F_1$ and $F_2$ be two fractional factorial designs with two
levels and let $S_1$ and $S_2$ be their corresponding state
polyhedra of $I(F_1),I(F_2)$. Then the nesting of state polyhedra
$S_1\subset S_2$ implies nesting of algebraic fans $\mathcal
L_a(F_1)\subset\mathcal L_a(F_2)$.
\end{lemma}

The following example is based upon Lemma \ref{lem_2k} and presents
an interesting relation between resolution and identifiability. That
is, bigger resolution points to more models in the algebraic fan.

\begin{example}\label{ex_resol}{\normalfont Let $F_1$ and $F_2$ be
the $2_\textnormal{IV}^{4-1}$ and $2_\textnormal{III}^{4-1}$
fractional fractional designs with eight runs in four factors and
respective generators $x_1x_2x_3x_4-1=0$ and $x_1x_2x_3-1=0$. The
subindices III, IV refer to the \textit{resolution} of the fraction,
see \cite{BH1961I,BH1961II}\index{design!resolution}. Their
corresponding state polyhedra are nested, i.e. $S( F_2)\subset S(
F_1)$ and by direct computation we confirm that the algebraic fans
are also nested. The algebraic fan $\mathcal L_a(F_2)$ has four
models, while $\mathcal L_a(F_1)$ includes $12$ elements.
}\end{example}

For fractional factorial designs, the estimation of interactions in
a design was related to the resolution of the design through the
property termed \textit{hidden projection}, see
\cite{EK2006,WW1995}. We conjecture the nesting of algebraic fans of
two designs $2^{k-p}$ with different resolution. However, exploiting
this nesting property of fans to compare designs using {\em
aberration} might need additional considerations.

\begin{example}\label{ex_aberr}{\normalfont Let $F_1,F_2$ be the fractions
 $2_\textnormal{IV}^{7-2}$ given by generators
$x_6-x_1x_2x_3=0,x_7-x_2x_3x_4=0$ and 
$x_6-x_1x_2x_3x_4=0,x_7-x_1x_2x_3x_5=0$ 
respectively. Although both fractions have the same resolution, the
fraction $F_2$ corresponds to a \textit{minimum aberration design}
using the definition of \cite{FH1980}. The state polyhedron $S(F_1)$
has $133$ vertices while $S(F_2)$ has $1708$. There is no nesting of
the state polyhedra and $\mathcal L_a(F_1)\cap\mathcal
L_a(F_2)\neq\emptyset$. }\end{example}

A proposal to compare two designs $D_1,D_2$ of the same size through
their state polytopes is to map the vertices of the state polytopes
$S(D_1),S(D_2)$ with a function $f:\mathbb R^d\to\mathbb R$. In this
way the state polytopes of $D_1$ and $D_2$ are compared by the
univariate projections of their vertices. We propose a weighted sum
of the vertex coordinates
\begin{equation}\label{eq_compsp}f(v_1,\ldots,v_d)=\sum_{i=1}^dw_iv_i,\end{equation}
with positive weights $w_i>0$. We use $w_i=1$ for $i=1,\ldots,d$ and
thus Equation (\ref{eq_compsp}) allows for direct comparison of
designs based on the distribution of total degrees for models in the
algebraic fan.

\begin{figure}[h]  
\begin{center}
\psset{unit=5.0mm,linewidth=0.8pt}
 \begin{pspicture}(58,0)(80,6)
\pspolygon[linewidth=0.5pt,fillstyle=vlines,hatchangle=5,hatchwidth=0.5pt,hatchsep=3pt,hatchcolor=black](60.5,0)(60.5,0.0468384074941452)(61.5,0.0468384074941452)(61.5,0)
\pspolygon[linewidth=0.5pt,fillstyle=vlines,hatchangle=5,hatchwidth=0.5pt,hatchsep=3pt,hatchcolor=black](61.5,0)(61.5,0.304449648711944)(62.5,0.304449648711944)(62.5,0)
\pspolygon[linewidth=0.5pt,fillstyle=vlines,hatchangle=5,hatchwidth=0.5pt,hatchsep=3pt,hatchcolor=black](62.5,0)(62.5,0.63231850117096)(63.5,0.63231850117096)(63.5,0)
\pspolygon[linewidth=0.5pt,fillstyle=vlines,hatchangle=5,hatchwidth=0.5pt,hatchsep=3pt,hatchcolor=black](63.5,0)(63.5,0.725995316159251)(64.5,0.725995316159251)(64.5,0)
\pspolygon[linewidth=0.5pt,fillstyle=vlines,hatchangle=5,hatchwidth=0.5pt,hatchsep=3pt,hatchcolor=black](64.5,0)(64.5,1.28805620608899)(65.5,1.28805620608899)(65.5,0)
\pspolygon[linewidth=0.5pt,fillstyle=vlines,hatchangle=5,hatchwidth=0.5pt,hatchsep=3pt,hatchcolor=black](65.5,0)(65.5,1.56908665105386)(66.5,1.56908665105386)(66.5,0)
\pspolygon[linewidth=0.5pt,fillstyle=vlines,hatchangle=5,hatchwidth=0.5pt,hatchsep=3pt,hatchcolor=black](66.5,0)(66.5,1.1943793911007)(67.5,1.1943793911007)(67.5,0)
\pspolygon[linewidth=0.5pt,fillstyle=vlines,hatchangle=5,hatchwidth=0.5pt,hatchsep=3pt,hatchcolor=black](67.5,0)(67.5,1.99063231850117)(68.5,1.99063231850117)(68.5,0)
\pspolygon[linewidth=0.5pt,fillstyle=vlines,hatchangle=5,hatchwidth=0.5pt,hatchsep=3pt,hatchcolor=black](68.5,0)(68.5,0.351288056206089)(69.5,0.351288056206089)(69.5,0)
\pspolygon[linewidth=0.5pt,fillstyle=vlines,hatchangle=5,hatchwidth=0.5pt,hatchsep=3pt,hatchcolor=black](69.5,0)(69.5,0.796252927400468)(70.5,0.796252927400468)(70.5,0)
\pspolygon[linewidth=0.5pt,fillstyle=vlines,hatchangle=5,hatchwidth=0.5pt,hatchsep=3pt,hatchcolor=black](70.5,0)(70.5,0.0468384074941452)(71.5,0.0468384074941452)(71.5,0)
\pspolygon[linewidth=0.5pt,fillstyle=vlines,hatchangle=5,hatchwidth=0.5pt,hatchsep=3pt,hatchcolor=black](71.5,0)(71.5,0.843091334894614)(72.5,0.843091334894614)(72.5,0)
\pspolygon[linewidth=0.5pt,fillstyle=vlines,hatchangle=5,hatchwidth=0.5pt,hatchsep=3pt,hatchcolor=black](73.5,0)(73.5,0.117096018735363)(74.5,0.117096018735363)(74.5,0)
\pspolygon[linewidth=0.5pt,fillstyle=vlines,hatchangle=5,hatchwidth=0.5pt,hatchsep=3pt,hatchcolor=black](79.5,0)(79.5,0.0936768149882904)(80.5,0.0936768149882904)(80.5,0)
\pspolygon[linewidth=0.5pt,fillstyle=vlines,hatchangle=5,hatchwidth=0.5pt,hatchsep=3pt,hatchcolor=black](118.5,0)(118.5,0)(119.5,0)(119.5,0)
\pspolygon[linewidth=0.5pt,fillstyle=vlines,hatchangle=105,hatchwidth=0.45pt,hatchsep=6pt,hatchcolor=red](60.5,0)(60.5,0)(61.5,0)(61.5,0)
\pspolygon[linewidth=0.5pt,fillstyle=vlines,hatchangle=105,hatchwidth=0.45pt,hatchsep=6pt,hatchcolor=red](61.5,0)(61.5,0)(62.5,0)(62.5,0)
\pspolygon[linewidth=0.5pt,fillstyle=vlines,hatchangle=105,hatchwidth=0.45pt,hatchsep=6pt,hatchcolor=red](62.5,0)(62.5,0)(63.5,0)(63.5,0)
\pspolygon[linewidth=0.5pt,fillstyle=vlines,hatchangle=105,hatchwidth=0.45pt,hatchsep=6pt,hatchcolor=red](63.5,0)(63.5,0)(64.5,0)(64.5,0)
\pspolygon[linewidth=0.5pt,fillstyle=vlines,hatchangle=105,hatchwidth=0.45pt,hatchsep=6pt,hatchcolor=red](64.5,0)(64.5,0)(65.5,0)(65.5,0)
\pspolygon[linewidth=0.5pt,fillstyle=vlines,hatchangle=105,hatchwidth=0.45pt,hatchsep=6pt,hatchcolor=red](65.5,0)(65.5,0)(66.5,0)(66.5,0)
\pspolygon[linewidth=0.5pt,fillstyle=vlines,hatchangle=105,hatchwidth=0.45pt,hatchsep=6pt,hatchcolor=red](66.5,0)(66.5,0)(67.5,0)(67.5,0)
\pspolygon[linewidth=0.5pt,fillstyle=vlines,hatchangle=105,hatchwidth=0.45pt,hatchsep=6pt,hatchcolor=red](67.5,0)(67.5,5.41353383458647)(68.5,5.41353383458647)(68.5,0)
\pspolygon[linewidth=0.5pt,fillstyle=vlines,hatchangle=105,hatchwidth=0.45pt,hatchsep=6pt,hatchcolor=red](68.5,0)(68.5,0)(69.5,0)(69.5,0)
\pspolygon[linewidth=0.5pt,fillstyle=vlines,hatchangle=105,hatchwidth=0.45pt,hatchsep=6pt,hatchcolor=red](69.5,0)(69.5,0)(70.5,0)(70.5,0)
\pspolygon[linewidth=0.5pt,fillstyle=vlines,hatchangle=105,hatchwidth=0.45pt,hatchsep=6pt,hatchcolor=red](70.5,0)(70.5,0)(71.5,0)(71.5,0)
\pspolygon[linewidth=0.5pt,fillstyle=vlines,hatchangle=105,hatchwidth=0.45pt,hatchsep=6pt,hatchcolor=red](71.5,0)(71.5,3.60902255639098)(72.5,3.60902255639098)(72.5,0)
\pspolygon[linewidth=0.5pt,fillstyle=vlines,hatchangle=105,hatchwidth=0.45pt,hatchsep=6pt,hatchcolor=red](73.5,0)(73.5,0)(74.5,0)(74.5,0)
\pspolygon[linewidth=0.5pt,fillstyle=vlines,hatchangle=105,hatchwidth=0.45pt,hatchsep=6pt,hatchcolor=red](79.5,0)(79.5,0.902255639097744)(80.5,0.902255639097744)(80.5,0)
\pspolygon[linewidth=0.5pt,fillstyle=vlines,hatchangle=105,hatchwidth=0.45pt,hatchsep=6pt,hatchcolor=red](118.5,0)(118.5,0.075187969924812)(119.5,0.075187969924812)(119.5,0)
\pspolygon[linewidth=0.5pt,fillstyle=vlines,hatchangle=-15,hatchwidth=0.85pt,hatchsep=8pt,hatchcolor=blue](58.5,0)(58.5,0.084)(57.5,0.084)(57.5,0)
\pspolygon[linewidth=0.5pt,fillstyle=vlines,hatchangle=-15,hatchwidth=0.85pt,hatchsep=8pt,hatchcolor=blue](59.5,0)(59.5,0.199)(58.5,0.199)(58.5,0)
\pspolygon[linewidth=0.5pt,fillstyle=vlines,hatchangle=-15,hatchwidth=0.85pt,hatchsep=8pt,hatchcolor=blue](60.5,0)(60.5,0.551)(59.5,0.551)(59.5,0)
\pspolygon[linewidth=0.5pt,fillstyle=vlines,hatchangle=-15,hatchwidth=0.85pt,hatchsep=8pt,hatchcolor=blue](61.5,0)(61.5,1.266)(60.5,1.266)(60.5,0)
\pspolygon[linewidth=0.5pt,fillstyle=vlines,hatchangle=-15,hatchwidth=0.85pt,hatchsep=8pt,hatchcolor=blue](62.5,0)(62.5,2.015)(61.5,2.015)(61.5,0)
\pspolygon[linewidth=0.5pt,fillstyle=vlines,hatchangle=-15,hatchwidth=0.85pt,hatchsep=8pt,hatchcolor=blue](63.5,0)(63.5,2.118)(62.5,2.118)(62.5,0)
\pspolygon[linewidth=0.5pt,fillstyle=vlines,hatchangle=-15,hatchwidth=0.85pt,hatchsep=8pt,hatchcolor=blue](64.5,0)(64.5,1.752)(63.5,1.752)(63.5,0)
\pspolygon[linewidth=0.5pt,fillstyle=vlines,hatchangle=-15,hatchwidth=0.85pt,hatchsep=8pt,hatchcolor=blue](65.5,0)(65.5,1.044)(64.5,1.044)(64.5,0)
\pspolygon[linewidth=0.5pt,fillstyle=vlines,hatchangle=-15,hatchwidth=0.85pt,hatchsep=8pt,hatchcolor=blue](66.5,0)(66.5,0.51)(65.5,0.51)(65.5,0)
\pspolygon[linewidth=0.5pt,fillstyle=vlines,hatchangle=-15,hatchwidth=0.85pt,hatchsep=8pt,hatchcolor=blue](67.5,0)(67.5,0.25)(66.5,0.25)(66.5,0)
\pspolygon[linewidth=0.5pt,fillstyle=vlines,hatchangle=-15,hatchwidth=0.85pt,hatchsep=8pt,hatchcolor=blue](68.5,0)(68.5,0.124)(67.5,0.124)(67.5,0)
\pspolygon[linewidth=0.5pt,fillstyle=vlines,hatchangle=-15,hatchwidth=0.85pt,hatchsep=8pt,hatchcolor=blue](69.5,0)(69.5,0.058)(68.5,0.058)(68.5,0)
\pspolygon[linewidth=0.5pt,fillstyle=vlines,hatchangle=-15,hatchwidth=0.85pt,hatchsep=8pt,hatchcolor=blue](70.5,0)(70.5,0.021)(69.5,0.021)(69.5,0)
\pspolygon[linewidth=0.5pt,fillstyle=vlines,hatchangle=-15,hatchwidth=0.85pt,hatchsep=8pt,hatchcolor=blue](71.5,0)(71.5,0.005)(70.5,0.005)(70.5,0)
\pspolygon[linewidth=0.5pt,fillstyle=vlines,hatchangle=-15,hatchwidth=0.85pt,hatchsep=8pt,hatchcolor=blue](72.5,0)(72.5,0.001)(71.5,0.001)(71.5,0)
\pspolygon[linewidth=0.5pt,fillstyle=vlines,hatchangle=-15,hatchwidth=0.85pt,hatchsep=8pt,hatchcolor=blue](73.5,0)(73.5,0)(72.5,0)(72.5,0)
\pspolygon[linewidth=0.5pt,fillstyle=vlines,hatchangle=-15,hatchwidth=0.85pt,hatchsep=8pt,hatchcolor=blue](74.5,0)(74.5,0)(73.5,0)(73.5,0)
\pspolygon[linewidth=0.5pt,fillstyle=vlines,hatchangle=-15,hatchwidth=0.85pt,hatchsep=8pt,hatchcolor=blue](80.5,0)(80.5,0)(79.5,0)(79.5,0)

\psset{linewidth=0.5pt}
\psline{<->}(57,6)(57,0)(81,0)
 \qline(56.8,1)(57,1) \qline(56.8,2)(57,2)
\qline(56.8,3)(57,3) \qline(56.8,4)(57,4) \qline(56.8,5)(57,5)
\psline{->}(65.5,2.5)(66,1.6) \rput(65.5,2.8){$F_2$}
\psline{->}(69.5,4.4)(68.7,4.4) \rput(70,4.4){$F_1$}
\psline{->}(60.6,2.4)(61.0,1.35) \rput(60.7,2.8){$F_3$}
\qline(60,0)(60,-0.2) \qline(62,0)(62,-0.2) \qline(64,0)(64,-0.2)
\qline(66,0)(66,-0.2) \qline(68,0)(68,-0.2) \qline(70,0)(70,-0.2)
\qline(72,0)(72,-0.2) \qline(74,0)(74,-0.2) \qline(76,0)(76,-0.2)
\qline(78,0)(78,-0.2) \qline(80,0)(80,-0.2)
\rput(60,-0.5){\tiny{$60$}} \rput(70,-0.5){\tiny{$70$}}
\rput(80,-0.5){\tiny{$80$}}
\rput(56.25,1){\tiny{$10\%$}} \rput(56.25,2){\tiny{$20\%$}}
\rput(56.25,3){\tiny{$30\%$}} \rput(56.25,4){\tiny{$40\%$}}
\rput(56.25,5){\tiny{$50\%$}}
\end{pspicture}

\caption{Histograms of relative frequencies for fractions $F_1$ and
$F_2$, see Example \ref{ex_aberr2}. We added $F_3$ of Example
\ref{ex_nonorth}.} \label{fig_histsp}
\end{center}
\end{figure}
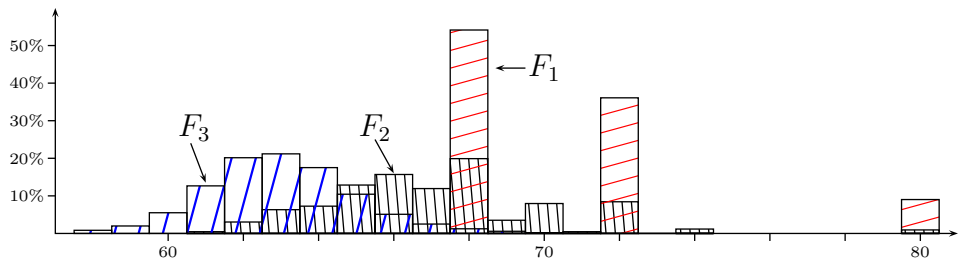

\begin{example}\label{ex_aberr2}{\normalfont (Continuation of Example \ref{ex_aberr})
We transform the vertices of the state polytopes for $F_1$ and $F_2$
using Equation (\ref{eq_compsp}). In Table \ref{Tab_comparison} in
Appendix B we summarize the results for each fraction as the
distribution of absolute and relative frequencies. Clearly, the
fraction $F_2$ with minimum aberration for generators identifies
models with a smaller total degree than that for $F_1$ and in that
sense it has smaller linear aberration. See Figure \ref{fig_histsp}
for a histogram of the relative frequencies for $F_1$ and $F_2$.
 }\end{example}

\section{Discussion}\label{sec_Di}

\subsection{Generalised concave aberration}

This paper is partly concerned with a problem of linear programming,
i.e. optimising a linear function $f:\mathbb R^d\to\mathbb R$ over a
convex polytope. We now discuss extensions of our work using other
types of aberration. When we consider concave aberration criteria,
some of our results still hold.

Consider any concave function $f:\mathbb{R}^d\to\mathbb{R}$. Now,
given a model $L$, define its aberration by
$$A(f,L):=
f \left(\sum_{\alpha\in L}\alpha_1,\dots,\sum_{\alpha\in
L}\alpha_d\right).$$ The linear aberration of Definition
\ref{def_ab} is the special case where $f$ is the following linear
(hence concave) function,
$$f:\mathbb{R}^d\; \longrightarrow\; \mathbb{R} $$
$$x=(x_1,\dots,x_d)\; \mapsto\; \frac{1}{n}\sum_{i=1}^d w_ix_i.$$


Since we only appealed to convexity, Theorem \ref{th_vertex_mini} is
valid when we replace $A(w,L)$ by the more general form $A(f,L)$.
That is to say, the set of lower vertices of the state polytope
(corresponding to models in the algebraic fan) contains the solution
to minimising any concave aberration function. This can be understood
as minimisation over a matroid, which was studied further 
in \cite{BLMORWW2008}.
 A further development is to consider
aberration $A(w,S(w))$ with respect to other distributions rather
than the uniform.

\subsection{Connection with aberration of Wu and others}
In the statistical literature, the word aberration has been used to
refer to properties of the generators for fractional factorial
designs, see \cite{CH1998,FH1980,WW2002}. A topic of future research
is to link minimal aberration of Definition \ref{def_ab} with the
traditional measure based on generators for a fractional factorial
design.

We conjecture that among the class of orthogonal fractions of $2^d$
designs there is some kind of correspondence between the minimal
linear aberration of this paper and minimum generator aberration of
Wu and others. If we select non-orthogonal fractions, the situation
is more complex, as the next example shows.

\begin{example}{\normalfont \label{ex_nonorth}
Let $F_3$ be the non-orthogonal fraction with size $n=32$ of a $2^7$
design given in Table \ref{Tab_F3} of Appendix B. We also consider
the designs $F_1$ and $F_2$ of Examples \ref{ex_aberr} and
\ref{ex_aberr2}. The three designs have the same size, but the
design $F_3$ cannot be compared with $F_1$ or $F_2$ in traditional
terms as it is not even orthogonal. However, we can compare the
designs based in the distribution of degrees in their algebraic
fans.

An interpolation as presented in Appendix B suggests that the
minimum degree of models identified by a generic design with
$n=32,d=7$ is $53.5\approx 54$. This number is a lower bound for the
total degree of models identified by designs $F_1,F_2$ and $F_3$. In
other words, the set of total degrees for models in algebraic fan of
$F_1,F_2$ and $F_3$ is lower bounded by 54, e.g. $54\leq
\min(\{\sum_{i=1}^d\bar\alpha_L:L\in\mathcal L_a(F_i)\})$ for
$i=1,2,3$.

Initial results show that
\begin{enumerate}\item[i)] the size of $\mathcal L_a(F_3)$ is much
longer (it has around $6\times10^5$ models) than that for designs
$F_1$ and $F_2$, see Table \ref{Tab_comparison} in Appendix B;
\item[ii)] the algebraic fans of $F_1$ and $F_2$ are not contained in
the algebraic fan of $F_3$, and
\item[iii)] the design $F_3$ identifies model of lower degree than $F_1$ or $F_2$ (indeed of
total degree $58$), and the bound $54$ is verified.
\end{enumerate}
It is clear that $F_3$ has smaller minimal linear aberration than
$F_1$ and $F_2$, see Figure \ref{fig_histsp}. We also note that the
histogram for $F_3$ presents more symmetry than that for $F_1$ and
$F_2$.}\end{example}

\section*{Acknowledgments}

The research of Shmuel Onn and Henry Wynn was partially supported by
the Joan and Reginald Coleman-Cohen Exchange Program during a stay
of Henry Wynn at the Technion-Israel Institute of Technology. Yael
Berstein was supported by an Irwin and Joan Jacobs Scholarship and
by a scholarship from the Graduate School of the Technion. Shmuel
Onn was also supported by the ISF: Israel Science Foundation. Henry
Wynn and Hugo Maruri-Aguilar were also supported by the Research
Councils UK (RCUK) Basic Technology grant ``Managing Uncertainty in
Complex Models".

\section*{Appendix A: Proof of Theorem \ref{th_apr_ab}} \label{app_proof}

\begin{proof} The proof is basically the minimisation over the first orthant
of $\sum_{i=1}^d{w_ix_i}$ subject to the constraint
$\prod_{i=1}^d(x_i+a)=b$. The problem is solved by a change of
coordinates to $x'_i=x_i+a$ for $i=1,\ldots,d$. We minimise
$\sum_{i=1}^d{w_ix'_i}$ subject to $\prod_{i=1}^dx'_i-b=0$ . Using
standard optimization tools, we form the Lagrange multiplier
\[L(x',\lambda)=\sum_{i=1}^d{w_ix'_i}-\lambda\left(\prod_{i=1}^dx'_i-b\right)\]
and then solve the system of equations $\nabla L(x',\lambda)= 0,
\frac{\partial L(x,\lambda)}{\partial\lambda}=0$. The solution
vector is $x^{*'}=(x_1^{*'},\ldots,x_d^{*'})$ where
\begin{equation*}\label{eq_minimum}x_i^{*'}=b^{1/d}\frac{\prod_{i=1}^d w_i^{1/d}}{w_i}.\end{equation*} The
convexity of the functions $\sum_{i=1}^d{w_ix_i}$ and
$\prod_{i=1}^dx_i=b$ over the first orthant guarantees that $x^{*'}$
is indeed the minimum. The aberration for this minimal point is
\[\sum_{i=1}^d{w_ix_i^*}=db^{1/d}g(w).\] Finally we note that
$x^*_i=x^{*'}_i-a$ and compute the aberration using $x^*_i$,
achieving the approximate aberration $\tilde A$ of Equation
(\ref{eq_minab}).
\end{proof}

We remark that for a fixed $w$, $x^*_i$ serves as an approximation
to the centroid of the corresponding corner cut model and therefore
$\tilde A$ is an approximation to $A^*(w,n)$. Although the
approximate aberration $\tilde A$ does not depend on the actual
corner cut identified by $L$, the minimal aberration $A^*(w,n)$ does
depend on it.
 If $L$ is the
corner cut directed by $w$, the practical validity of the
approximate aberration $\tilde A$ relies on $x^*_i$ being close
enough to $\frac{1}{n}\sum_{\alpha\in L}\alpha_i$. This closeness
depends ultimately on $a,b$. See Appendix \ref{ap_computeab} for a
proposal to compute $a,b$.

\section*{Appendix B: Computing values $a,b$ for the approximate corner cut
polytope}\label{ap_computeab}


In Section \ref{sec_ap_sp} we proposed the continuous function of
Equation (\ref{ec_aprox}) to approximate the corner cut polytope
(which is piecewise linear surface). In this section we discuss on
the selection of the values $a,b$ so that the approximation is good
enough. In general, the values $a,b$ will depend on the number of
dimensions $d$ and number of points in the design $n$. However, for
fixed $d$, the approximation will be coarse for small values of $n$.

For our approximation we use the following properties of the corner
cut polytope, which have been studied as well in \cite{IM2003} and
\cite{OS1999}.
\begin{lemma}\label{lem_ccpoly}
The corner cut polytope satisfies the following properties.
\begin{enumerate}
\item[i)] \label{lem1}The intersection of the corner cut polytope with the axes
occurs at the point ${n\choose 2}$.
\item[ii)] \label{lem2}When for $k\geq 1$, the sample size $n$ satisfies \begin{equation}\label{eq_nk}n={k+d-1\choose d}\end{equation} then the
corner cut polytope is pointed.
\end{enumerate}
\end{lemma}

\begin{proof}
\begin{enumerate}
\item[i)] The intersection is the
the sum of exponents for any marginal model of the form
$\{1,x_i,x_i^2,\ldots,x_i^{n-1}\}$. Therefore the intersection must
occur at $\sum_{i=0}^{n-1}i={n\choose 2}$. \item[ii)] The corner cut
polytope is pointed when the sample size is the same as the size of
a model of total degree $k-1$, that is,
 there are ${d+1-j\choose j}$ terms of degree $j$ in
the model where $j=0,\ldots,k-1$. Therefore the sample size must be
$n=\sum_{j=0}^{k-1}{d+1-j\choose j}={k+d-1\choose d}$.
\end{enumerate}
\end{proof}

\begin{remark}When Equation (\ref{eq_nk}) is satisfied, the tip of the
pointed corner cut polytope has coordinates
$\alpha_L=\left({k+d-1\choose d+1},\ldots,{k+d-1\choose
d+1}\right)$.\end{remark}

We propose to force Equation (\ref{ec_aprox}) to satisfy the
condition of Item 1 in Lemma \ref{lem_ccpoly} and pass through the
tip point $\alpha_L$ for the model of total degree $k-1$. To
summarize, when sample size satisfies Equation (\ref{eq_nk}) then
$a,b$ must satisfy the following equations:
$$b=a^{d-1}\left(\frac{n-1}{2}+a\right)\mbox{ and } b=(c+a)^d,$$
where $c=\frac{1}{n}{k+d-1\choose d+1}$ is the scaled tip of the
corner cut poytope. When design size, $n$, is not of the form
$n={k+d-1\choose d}$ for some $k\geq 1$, we propose to interpolate
the value for $c$, the scaled tip of the polytope, that is to solve
Equation (\ref{eq_nk}) for $k$ and interpolate the corresponding tip
with $\frac{1}{n}{k+d-1\choose d+1}$.

\begin{figure}[h]  
\begin{center}
\psset{unit=3.80mm,linewidth=0.8pt}
 \begin{pspicture}(0,-2)(22,22)
     \put(0,0){}

\psset{linewidth=0.25pt}
\psline(0,21)(1,15)(2,11)(4,7)(7,4)(11,2)(15,1)(21,0)
\psdots[dotsize=0.15pt,dotstyle=*](0,21)(1,15)(2,11)(4,7)(7,4)(11,2)(15,1)(21,0)
\psset{linewidth=0.5pt}
\psline[linestyle=solid,linewidth=0.15pt]{-}(0,0)(6,9)
\rput(2.1,4.6){$w$}
\psline[linestyle=solid]{->}(0,0)(2.8, 4.2)
\psset{linewidth=0.5pt}
\psline[linestyle=dashed]{-}(7.110965452,3.836139499)(3.958515293, 5.937772939)
\psline[linestyle=dashed]{->}(0,0)(7.110965452,3.836139499)
\psset{linewidth=1pt}
\psline[linestyle=dotted]{-}(7,4)(4,6)
\psline[linestyle=dotted]{->}(0,0)(7,4)
\psset{linewidth=0.5pt}
\psline[linestyle=dashed,linewidth=0.25pt](0,21)(0.295774647887324,18.6692632996307)
(0.591549295774648,16.7556884489958)(0.887323943661972,15.1564777473601)
(1.1830985915493,13.8000451840891)(1.47887323943662,12.6350064464558)
(1.77464788732394,11.6235223502531)(2.07042253521127,10.7371115905248)
(2.36619718309859,9.95392635015022)(2.66197183098592,9.25692733167246)
(2.95774647887324,8.63262985738038)(3.25352112676056,8.0702228901776)
(3.54929577464789,7.56093769130891)(3.84507042253521,7.09758730923431)
(4.14084507042254,6.67422529851105)(4.43661971830986,6.2858891439377)
(4.73239436619718,5.92840483665912)(5.02816901408451,5.59823624785275)
(5.32394366197183,5.29236776028493)(5.61971830985915,5.00821189469427)
(5.91549295774648,4.74353593380409)(6.2112676056338,4.49640313678325)
(6.50704225352113,4.2651252679539)(6.80281690140845,4.0482239781593)
(7.09859154929577,3.84439917081861)(7.3943661971831,3.65250292197729)
(7.69014084507042,3.47151784905707)(7.98591549295775,3.30053906744746)
(8.28169014084507,3.13875905933987)(8.57746478873239,2.98545492078731)
(8.87323943661972,2.83997756202757)(9.16901408450704,2.70174252073132)
(9.46478873239437,2.57022211396324)(9.76056338028169,2.44493870665912)
(10.056338028169,2.32545891559427)(10.3521126760563,2.21138860060196)
(10.6478873239437,2.10236852105334)(10.943661971831,1.99807055674537)
(11.2394366197183,1.89819440944653)(11.5352112676056,1.80246471525782)
(11.830985915493,1.71062850930937)(12.1267605633803,1.6224529936367)
(12.4225352112676,1.53772356676564)(12.7183098591549,1.45624207989351)
(13.0140845070423,1.37782528983672)(13.3098591549296,1.30230348332028)
(13.6056338028169,1.22951925087076)(13.9014084507042,1.15932639167037)
(14.1971830985915,1.09158893333819)(14.4929577464789,1.02618025280964)
(14.7887323943662,0.962982286354718)(15.0845070422535,0.901884818365692)
(15.3802816901408,0.842784839900806)(15.6760563380282,0.78558596913042)
(15.9718309859155,0.730197926826519)(16.2676056338028,0.676536060891546)
(16.5633802816901,0.624520914659542)(16.8591549295775,0.574077834339293)
(17.1549295774648,0.525136611520619)(17.4507042253521,0.47763115714361)
(17.7464788732394,0.431499203746938)(18.0422535211268,0.386682033174373)
(18.3380281690141,0.343124227235675)(18.6338028169014,0.300773439095578)
(18.9295774647887,0.259580183407976)(19.2253521126761,0.219497643426283)
(19.5211267605634,0.180481493509187)(19.8169014084507,0.14248973560706)
(20.112676056338,0.105482548460914)(20.4084507042254,0.0694221483755739)
(20.7042253521127,0.0342726605437815)(21,0)
\psset{linewidth=0.3pt,linestyle=solid}

\rput(-0.5,-0.5){$0$}
\rput(21,-0.5){$3$} \rput(14,-0.5){$2$} \rput(7,-0.5){$1$}
\rput(-0.5,21){$3$} \rput(-0.5,14){$2$} \rput(-0.5,7){$1$}

 \psline{->}(0,0)(22,0)
\psline(1,0)(1,-0.15)\psline(2,0)(2,-0.15)\psline(3,0)(3,-0.15)
\psline(4,0)(4,-0.15)\psline(5,0)(5,-0.15)
\psline(6,0)(6,-0.15)\psline(7,0)(7,-0.15)\psline(8,0)(8,-0.15)
\psline(9,0)(9,-0.15)\psline(10,0)(10,-0.15)
\psline(11,0)(11,-0.15)\psline(12,0)(12,-0.05)\psline(13,0)(13,-0.05)
\psline(14,0)(14,-0.15)\psline(15,0)(15,-0.05)
\psline(16,0)(16,-0.15)\psline(17,0)(17,-0.05)\psline(18,0)(18,-0.05)
\psline(19,0)(19,-0.15)\psline(20,0)(20,-0.05)\psline(21,0)(21,-0.05)

\psline{->}(0,0)(0,22)
\psline(0,1)(-0.15,1)\psline(0,2)(-0.15,2)\psline(0,3)(-0.15,3)
\psline(0,4)(-0.15,4)\psline(0,5)(-0.15,5)
\psline(0,6)(-0.15,6)\psline(0,7)(-0.15,7)\psline(0,8)(-0.15,8)
\psline(0,9)(-0.15,9)\psline(0,10)(-0.15,10)
\psline(0,11)(-0.15,11)\psline(0,12)(-0.15,12)\psline(0,13)(-0.15,13)
\psline(0,14)(-0.15,14)\psline(0,15)(-0.15,15)
\psline(0,16)(-0.15,16)\psline(0,17)(-0.15,17)\psline(0,18)(-0.15,18)
\psline(0,19)(-0.15,19)\psline(0,20)(-0.15,20)\psline(0,21)(-0.15,21)

\end{pspicture}

\caption{Minimal aberration using the corner cut
polytope. The corner cut polytope is the piecewise linear solid curve, while the
approximation is the dashed curve.
The minimal aberration is the projection over the direction of $w$ of the
vertex (using dotted line), and an approximate value uses Equation (\ref{ec_aprox})
(dashed line).}  \label{fig_ab5}
\end{center}
\end{figure}
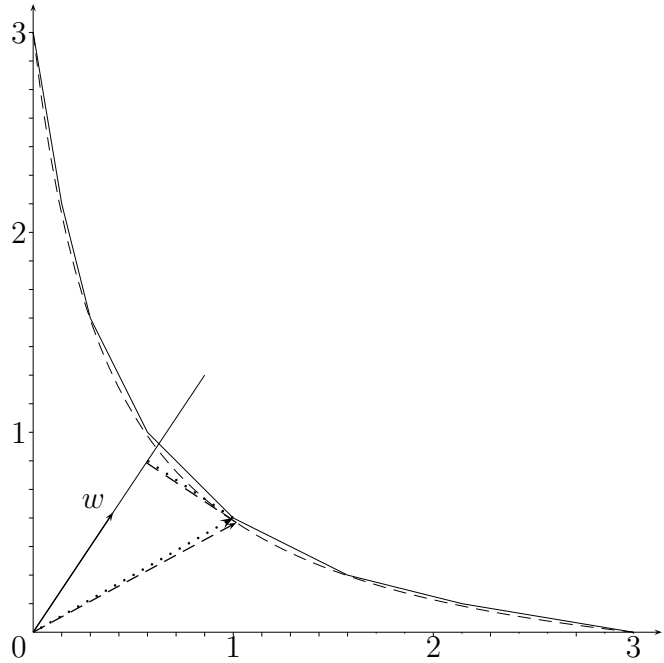

For two dimensions ($d=2$) by interpolation and solving the two
conditions above we obtain the following formul{\ae} for $a,b$ in
terms of $n$:
$$a=\frac{5-3\sqrt{1+8n}+4n}{3(3-2\sqrt{1+8n}+3n)},\; b=a\left(\frac{n-1}{2}+a\right).$$
See Figure \ref{fig_ab5} for a depiction of the corner cut polytope
and the approximate curve for $d=2,n=7$. This interpolation is
difficult for $d>2$ and we have to rely on approximations. The
following formul{\ae} are rough approximations for $a,b$ obtained by
truncation of the binomial expansions
$$a\approx\left(\frac{2d!n}{(d+1)^d(n-1)}\right)^{\frac{1}{d-1}}\mbox{, }b=a^{d-1}\left(\frac{n-1}{2}+a\right)\approx\frac{d!n}{(d+1)^d}.$$

\begin{table}[h]
\begin{center}\begin{tiny}
\begin{tabular}{rrrrrrr}
{\begin{tabular}{l}
Total\\degree\end{tabular}}&AF $F_1$& AF $F_2$& AF $F_3$&RF $F_1$&RF $F_2$ &RF $F_3$\\
\hline
58& -& -&2290& -& -&0.84\\
59& -& -&5437& -& -&1.99\\
60& -& -&15036& -& -&5.51\\
61& -&8&34574& -&0.47&12.66\\
62& -&52&55025& -&3.04&20.15\\
63& -&108&57848& -&6.32&21.18\\
64& -&124&47851& -&7.26&17.52\\
65& -&220&28511& -&12.88&10.44\\
66& -&268&13928& -&15.7&5.1\\
67& -&204&6837& -&11.94&2.5\\
68&72&340&3378&54.14&19.91&1.24\\
69& -&60&1596& -&3.51&0.58\\
70& -&136&567& -&7.96&0.21\\
71& -&8&140& -&0.47&0.05\\
72&48&144&33&36.09&8.43&0.01\\
73& -& -&12& -& -&0.00\\
74& -&20&5& -&1.17&0.00\\
80&12&16& -&9.02&0.94&0.00\\
83& -& -&1& -& -&0.00\\
85& -& -&1& -& -&0.00\\
119&1& -& -&0.75& -& -\\
 \hline
Total& 133&1708&273071&100.00&100.00&100.00\\
\end{tabular}\end{tiny}\caption{Absolute (AF) and relative (RF) frequencies of total
degrees for models identified by fractions $F_1$ and $F_2$ of
Example \ref{ex_aberr2} and $F_3$ of Example \ref{ex_nonorth}. The
symbol - represents zero.} \label{Tab_comparison}
\end{center}
\end{table}


\begin{table}[h]
\begin{center}\begin{tiny}
\begin{tabular}{ccccccc}
$x_1$&$x_2$&$x_3$&$x_4$&$x_5$&$x_6$&$x_7$\\ \hline
+&+&+&+&-&-&+\\
+&-&+&-&-&+&+\\
+&-&+&+&-&+&-\\
+&+&+&-&+&+&-\\
+&+&-&-&-&-&+\\
+&-&+&+&-&-&+\\
+&-&-&-&+&+&+\\
+&-&-&+&-&-&+\\
-&+&+&-&+&-&-\\
+&-&-&+&-&+&-\\
+&-&+&-&+&-&-\\
-&+&+&+&-&-&+\\
-&+&+&+&+&-&-\\
+&-&-&+&+&+&-\\
-&-&-&-&-&-&-\\
+&-&-&-&+&-&-\\
-&+&+&+&+&-&+\\
-&-&+&+&-&+&-\\
+&-&-&-&-&+&-\\
-&-&-&-&+&+&+\\
-&-&+&-&-&+&+\\
+&-&+&-&+&+&-\\
-&+&+&-&+&+&-\\
-&+&-&-&-&+&+\\
-&-&-&+&+&-&+\\
+&+&-&-&+&+&+\\
+&+&+&+&-&+&+\\
-&-&-&-&-&-&+\\
-&-&+&-&+&-&+\\
+&+&-&-&+&-&+\\
-&-&-&-&-&+&+\\
+&+&-&-&-&-&-\\
\end{tabular}\end{tiny}
\caption{Design $F_3$ of Example \ref{ex_nonorth}. The signs $+$ and
$-$ correspond to $+1$ and $-1$.}\label{Tab_F3}
\end{center}
\end{table}

\bibliographystyle{apalike} 
\bibliography{notes}
\end{document}